\documentclass[12p]{article}

\begin{document}

\title{Korteweg-de Vries hierarchy and related completely integrable systems:\\
I. Algebro-geometrical approach \\
Perprint TH-98/4, 1998, INRNE,BAS, Sofia, Bulgaria}
\author{N.~A.~Kostov \\
Institute of Electronics\\
Bulgarian Academy of Sciences\\
Blvd. Tsarigradsko shosse 72, Sofia 1784, Bulgaria}

\maketitle

\begin{abstract}
We consider complementary dynamical systems related to stationary Korteweg-
de Vries hierarchy of equations. A general approach for finding elliptic
solutions is given. The solutions are expressed in terms of Novikov
polynomials in general quasi-periodic case. For periodic case
these polynomials coincide with Hermite and Lam\'e polynomials. As byproduct
we derive $2\times 2$ matrix Lax representation for  Rosochatius-Wojciechowski,
Rosochatius system, second flow of stationary nonlinear vector Schr\"{o}dinger
equations and complex Neumann systems.
\end{abstract}



\section{Introduction}

\setcounter{equation}{0} The problem of integrability and nonintegrability
of Hamiltonian system with $g$ degrees of freedom has been a subject of
considerable interest for many years. Recently remarkable progress has been
achieved in connection of study of stationary soliton type equations. The
method of restricted flows was introduced in \cite{Mos1,20,Mos2} as a
non-linearization of the Korteweg-de Vries (KdV) spectral problem and was
generalized in \cite{5,2}. The coupled Neumann system, the Neumann system,
the Garnier type systems, the Rosochatius-Wojciechowski, Rosochatius and the
H\'enon-Heiles type
systems are examples of this type. In this relation important results about
the algebro-geometrical interpretaion of these systems are obtained in \cite
{F}. General approach to completely integrable dynamical systems of Neumann
type is discussed in \cite{S, AHP, AHH}. Quasi-periodic solutions and
spectral interpretation using both algebro-geometrical and spectral methods
are given in \cite{K}. Quasi-periodic solutions ($g=2$) and
periodic solutions associated to Lam\'e and Treibich-Verdier potentials are
obtained in \cite{EK, CEEK}. New method of constructing elliptic finite-gap
solutions of the stationary KdV hierarchy, based on a theorem due to Picard,
is proposed in \cite{GW95a,GW95b,GW95c,GW95d,GW96,GW98}.

In this paper we are concerned with the following different approaches

\begin{itemize}
\item  integrable dynamical systems related to Hill's equation in the case
of finite-gap potential \cite{20, AM80, F, AHP, AHH, S, K, CC}

\item  method of stationary flows and restricted flows \cite{1,2}

\item  method of non-linearization of the KdV spectral problem \cite{5}

\item  method of separation of variables \cite{KKM, KRT, EEKT}

\item  algebro-geometrical construction \cite{F,S,K}.
\end{itemize}

We give unified construction based on algebro-geometrical approach. New
solutions in terms of Novikov and Gelfand-Dickey (GD) polynomials are given
in explicit form.

The paper is organized as follows. In section 2 we construct the stationary
flows associated to the KdV hierarchy strictly following \cite{1,7,8,9}.
Some maps between completely integrable dynamical systems are presented. In
section 3 we analyse the relation between restricted KdV flows and Garnier
type systems following \cite{8}. The map between stationary and restricted
KdV flows are given in section 4 \cite{8}. The list of known dynamical
systems related to stationary KdV equations are presented in section 3.

In section 4 we formulate Baker-Akhiezer function approach from
algebro-geometrical and spectral point of view respectively. New solutions
of integrable dynamical system associated to stationary KdV equations in
terms of Novikov polynomials are presented . In
particular these polynomials associated to Lam\'e potentials coincide with
Hermite polynomials and Lam\'e polynomials. In section 5 ($2\times 2$) Lax
representations of Garnier, Rosochtius I and stationary second flow of
vector nonlinear Schr\"{o}dinger equation are given. Following \cite
{EEKT,KRT,T1} $r$-matrix approach of these systems is discussed.

\section{KdV hierarchy and Gelfand-Dickey polynomials}

\setcounter{equation}{0} In this section we follow the geometrical
construction of \cite{1,7,8,9,v98} with only little changes of signs and
different spectral parameter. Let $M$ be a bi-Hamiltonian manifold: if the
associated Poisson operator $P^{\lambda}:=P_{1}+4\lambda P_{0}$ admits as a
Casimir a formal Laurent series $h(\lambda)$
\begin{equation}
h(\lambda):=\sum_{j\geq0}\,h_{j}\lambda^{-j}
\end{equation}
then $h_{0}$ is a Casimir of $P_{0}$ and the coefficients $h_{j}$ ($j\geq0$)
are the Hamiltonian functions of a hierarchy of bi-Hamiltonian vector fields
$X_{j}$:
\begin{equation}
X_{j}=P_{1}dh_{j}=P_{0}dh_{j+1}, \quad (j\geq0)  \label{VecF}
\end{equation}
At any point $u\in M$, the bi-Hamiltonian flows are given by $\frac{d u}{%
dt_{j}} =X_{j}(u)$, $t_{j}$ being the evolution parameter of the $j$th flow.
The vector field (\ref{VecF}) are Hamiltonian also with respect to the
Poisson operator $P^{\lambda}$. In fact the recursion relation (\ref{VecF})
can be written as
\begin{equation}
X_{j}=P^{\lambda}\,dh^{(j)}(\lambda),\quad h^{(j)}(\lambda):=(\lambda^{j}
h(\lambda))_{+}  \label{X}
\end{equation}
where the index $+$ means the projection of a Laurent series onto the purely
polynomial part.

Let $M$ be the algebra of polynomials in $u, u_{x},u_{xx},\ldots$ ($u=u(x)$
is a $C^{\infty}$ function of $x$ and the subscript $x$ means the derivative
with respect to $x$), and le $P_{0}$ and $P_{1}$ be the two Poisson
operators of the KdV hierarchy \cite{11}:
\begin{equation}
P_{0}:=\frac{d}{dx}, \qquad P_{1}:=\frac{d^{3}}{dx^{3}} - 4 u \frac{d}{dx} -
2 u_{x} .  \label{Opers}
\end{equation}
The gradients of the Casimirs of the associated Poisson operator $%
P^{\lambda} $ can be obtained searching the functions $v(\lambda):=\sum_{j%
\geq0}v_{j}\lambda^{-j}$ which are solutions of the following equation:
\begin{equation}
B^{\lambda}(v(\lambda),v(\lambda))=a(\lambda)  \label{GDg}
\end{equation}
where $a(\lambda)=\sum_{j\geq0}a_{j}\lambda^{-j}$, $a_{j}$ are constant
parameters and $B^{\lambda}$ is the bilinear function
\begin{equation}
B^{\lambda}(w_{1},w_{2}):=w_{1xx}w_{2}+w_{1}w_{2xx}-w_{1x}w_{2x}-
4(u-\lambda)w_{1}w_{2} .  \label{Brel}
\end{equation}
In fact $B^{\lambda}$ is related to the Poisson operator through the
relation
\begin{equation}
\frac{d}{dx}B^{\lambda}(w_{1},w_{2})=w_{1} P^{\lambda} w_{2} +w_{2}
P^{\lambda} w_{1} .
\end{equation}
Equation (\ref{GDg}) can be solved developing the left-hand side as a
Laurent series
\begin{equation}
B^{\lambda}(v(\lambda),v(\lambda)) = \sum_{k\geq -1} B_{k}\lambda^{-k}
\end{equation}
so that, for each $a(\lambda)$, it furnishes the coefficient of the solution
$v(\lambda)$ (unique up to a sign). The solution corresponding to $\bar{a}%
(\lambda) =-\lambda$ is the so-called basis solution $\bar{v}(\lambda)$; its
first coefficients are
\begin{eqnarray}
\bar{v}_{0}=1,\quad \bar{v}_{1}=1, \quad \bar{v}=2(u_{xx}+3 u^{2}), \\
\bar{v}_{3}=2(u_{xxxx}+5u_{x}^{2}+10u_{xx}u+10u^{3}) ,
\end{eqnarray}
and so on, namely the gradients of the first KdV Hamiltonians. In what
follows we shall consider also the function $v(\lambda)=c(\lambda)\bar{v}%
(\lambda)$, which is a solution of (\ref{GDg}) for
\begin{equation}
a(\lambda)=-\lambda c^{2}(\lambda), \quad c(\lambda)=1 + \sum_{j\geq
1}\,c_{j}\lambda^{-j} ,
\end{equation}
where the coefficients $c_{j}$ are free parameters. In this case the first
1-forms of the hierarchy are $v_{0}=1, v_{1}=\bar{v}_{1}+c_{1}, v_{2}=\bar{v}%
_{2}+c_{1}\bar{v}_{1}+c_{2}$, and so on.

The coefficient $B_{k}$ can be expressed through the GD polynomials. For
each Laurent series $v(\lambda)$ let us consider the functions $%
B^{(k)}(\lambda):=B^{\lambda}(v(\lambda),v^{(k)}(\lambda))$, where $%
v^{(k)}(\lambda):=\left(\lambda^{k}v(\lambda)\right)_{+}$; these functions
have the form
\begin{equation}
B^{(k)}(\lambda)=\lambda^{k+1}v_{0}^{2}+
\sum_{j=1}^{k-1}\lambda^{k-j}(p_{0j} +v_{0}v_{j+1}) +\sum_{j\geq
0}\lambda^{-j}p_{jk} .  \label{BB}
\end{equation}
It can be shown that
\begin{equation}
B_{-1}=-v_{0}^{2}, \quad B_{k}=p_{0k}-v_{0}v_{k+1} .
\end{equation}
Furthermore, if $v(\lambda)$ is a solution of (\ref{GDg}), the coefficients $%
p_{jk}$ in (\ref{BB}) are polynomials in $u$ and its $x$-derivatives. They
will be referred to as Gelfand-Dickey (GD) polynomials and the function $%
B^{\lambda}$ as their generating function.

The fundamental property of the GD polynomials, stemming from (\ref{GDg}), (%
\ref{BB}) is the following relation with the gradients $v_{j}=dh_{j}$ and
the bi-Hamiltonian vector field $X_{k}$:
\begin{equation}
\frac{d}{dx}p_{jk}=v_{j}X_{k} .
\end{equation}
We report some GD polynomials to be used in what follows $(v_{0}=1)$:
\begin{eqnarray}  \label{GD}
p_{00}&=&4 u - v_{1} ,  \nonumber \\
p_{01}&=&8 u v_{1}-v_{1}^{2}-v_{2}+2 v_{1xx} ,  \nonumber \\
p_{02}&=&4uv_{1}^{2}+8uv_{2}-2v_{1}v_{2}-v_{3}-v_{1x}^{2}+
2v_{1}v_{1xx}+2v_{2xx},  \nonumber \\
\\
p_{12}&=&8uv_{1}v_{2}-v_{2}^{2}+4uv_{3}-v_{1}v_{3}-v_{4}+
2v_{1x}v_{2x}+2v_{2}v_{1xx}+2v_{1}v_{2xx}+v_{3xx} ,  \nonumber \\
p_{kk}&=&2v_{kxx}v_{k}-v_{k}^{2}+4uv_{k}^{2} .  \nonumber
\end{eqnarray}
The GD polynomials corresponding to the basis solution $\bar{v}(\lambda)$
are the polynomials defined in \cite{1}, proposition 12.1.12.

\subsection{The method of stationary flows}

The method of stationary flows was developed in
order to reduce the flows of the KdV hierarchy onto the set $M_{g}$ of fixed
points of the $g$th flow $X_{g}$ of the hierarchy
\begin{equation}
M_{g}:=\left\{u|X_{g}(u,u_{x},\ldots,u^{(2g+1)})=0\right\}.
\end{equation}
As $M_{g}$ is odd-dimensional it can not be a symplectic manifold;
nevertheless we will show that it is a bi-Hamiltonian manifold; it will be
referred to as extended phase space. Moreover, $M_{g}$ is naturally
foliated, on account of (\ref{VecF}) and (\ref{Opers}), by a one-parameter
family of $2g$-dimensional submanifolds $S_{g}$ given by
\begin{equation}
S_{g}:=\left\{u|v_{g+1}(u,u_{x},\ldots,u^{(2g)}=c\right\}
\end{equation}
($c$ being a constant parameter), which are invariant manifolds with respect
to each vector field of the KdV hierarchy, due to the invariance of the
functions $v_{k}$. So $M_{g}$ can be parametrized naturally by $%
v_{1},\ldots, v_{g+1}$ and by their $x$-derivatives $v_{1x},\ldots,v_{gx}$.
We shall use these coordinates in what follows.

From the computational point of view, one proceeds as follows.

(i) Due to (\ref{X}) and (\ref{GDg}), the manifold $M_{g}$ is defined by the
solutions $u$ of the equation
\begin{equation}
B^{\lambda}(v(\lambda),v^{(g)}(\lambda))=\lambda^{g}a(\lambda)  \label{BL}
\end{equation}
where $v(\lambda)=\sum_{j=0}^{g}v_{j}\lambda^{-j}$, $a(\lambda)=%
\sum_{j=-1}^{2g} a_{j}\lambda^{-j}$. In particular, if $a(\lambda)=-\lambda
c^{2}(\lambda)$, $M_{g}$ is given by
\begin{equation}
M_{n}=\left\{u|\bar{X}_{n}+\sum_{j=1}^{n}c_{j}\bar{X}_{n-j} = 0\right\}
\end{equation}
i.e. by the solutions of the Lax-Novikov equations. Taking into acount (\ref
{BB}) and choosing $a_{-1}=-1$, by equation (\ref{BL}) the coefficients of $%
\lambda^{g+1}$ we get $v_{0}^{2}=1$; from now on we put $v_{0}=1$. Moreover,
equating the coefficients of the other powers of $\lambda$ we get the
following system:
\begin{equation}
p_{0k} - v_{k+1} =a_{k},\quad (k=0\ldots, n-1)\quad p_{jn}=a_{n+j},\,
(j=0,\ldots,n).  \label{PP0}
\end{equation}

(ii) In order to obtain the first Poisson tensor $P_{0}$, we eliminate $%
u=v_{1}/2+a_{0}/4$ from (\ref{PP0}) using the first equation ($k=0$) and we
extract the system of $n$ second-order ODEs in the $v_{j}$ ($j=1,\ldots,n$):
\begin{equation}
p_{0k} - v_{k+1} =a_{k},\quad (k=1\ldots, g-1)\quad p_{0g}=a_{g}  \label{P0}
\end{equation}
which will be referred as $P_{0}$-system. The remaining equations (\ref{PP0}%
) will furnish a set of $g$ independent integrals of motion. In order to
obtain a second Poisson structure, we consider the following system: ($P_{1}$%
-system)
\begin{equation}
p_{0k} - v_{k+1} =a_{k},\quad (k=1\ldots, g-1)\quad p_{gg}=a_{2g},
\label{P1}
\end{equation}
with $u$ as above.

(iii) The $P_{0}$-system (\ref{P0}) and the $P_{1}$-system (\ref{P1}) can be
written as canonical Hamiltonian systems
\begin{equation}
r_{kx}=\frac{\partial H_{g}^{(0)}}{\partial s_{k}}, \qquad s_{kx}=-\frac{%
\partial H_{g}^{(0)}}{\partial r_{k}},
\end{equation}
\begin{equation}
q_{kx}=\frac{\partial H_{g}^{(1)}}{\partial p_{k}}, \qquad p_{kx}=-\frac{%
\partial H_{g}^{(1)}}{\partial q_{k}}  \label{Deq}
\end{equation}
have $n$ integrals of motion given by
\begin{eqnarray}
&&K_{j}=-\frac{1}{8}p_{jg}|_{r}=a_{g+j}, (j=1,\ldots,g), \\
&&H_{j}=-\frac{1}{8}p_{jg}|_{x}=a_{g+j}, (j=0,\ldots,g-1),
\end{eqnarray}
Moreover, the map $\Phi:M_{g}\rightarrow M_{g}$ in the extended phase space
generates a second Poisson structure.

\section{Integrable dynamical systems related to hierarchy of stationary KdV
equations}

\setcounter{equation}{0} In the recent years, remarkable progress has been
achieved in the description of those quasi(periodic) potentials which belong
to a given spectrum. Many integrable systems of differential equations are
shown to be closely connected with Hill's equation in the case of a finite
gap potential. The coupled Neumann system, The Neumann system, and the
Rosochatius systems are examples of this type.

In this paragraph we are concerned with the following completely integrable
systems

The Garnier system
\begin{eqnarray}
\xi_{ixx} = \left( 2\sum_{j=1}^{g} \xi_{j} \eta_{j} +
\tilde{a}_{i}\right) \xi_{i}
, \quad \eta_{ixx} = \left( 2\sum_{j=1}^{g} \xi_{j} \eta_{j} +
\tilde{a}_{i}\right) \eta_{i} .  \label{Garn}
\end{eqnarray}
The $g$-dimensional anisotropic harmonic oscillator in radial quartic
potential, is obtained when $\xi_{i}=\eta_{i}$ $i=1,\ldots, g$
\begin{eqnarray}
\xi_{ixx} = \left( 2\sum_{j=1}^{g} \xi_{j}^2 + \tilde{a}_{i}\right) \xi_{i} .
\label{oscil}
\end{eqnarray}
Another interesting integrable system was proposed recently \cite{W}, we
call it the Rosochatius-Wojciechowski system. In our context, this system is
obtained by the Deift elimination procedure. Let
\begin{equation}
\xi_{i}=\psi_{i}\exp(\theta_{i}),\, \eta_{i}=\psi_{i}\exp(-\theta_{i}) ,\,
\sqrt{f}_{i} = \psi_{i}^{2}\theta_{ix} ,  \label{Deift}
\end{equation}
then equations (\ref{Garn}) transform to the Rosochatius-Wojciechowski system
\begin{eqnarray}
\psi_{ixx} = \left( 2\sum_{j=1}^{g} \psi_{j}^2 \right) \psi_{i}
+\tilde{a}_{i}\psi_{i}-f_{i}/\psi_{i}^3 , \,i=1,\ldots ,g  \label{RosI}
\end{eqnarray}
and the Hamiltonian is given by
\begin{eqnarray}
H=\sum_{j=1}^{g}\,\chi_{j}^{2} -
\left(\sum_{k=1}^{g}\,\psi_{k}^{2}\right)^{2} - \sum_{j=1}^{n}
\tilde{a}_{j}\psi_{j}^{2} -\sum_{j=1}^{g} \frac{f_{j}}{\psi_{j}^{2}} ,
\end{eqnarray}
where $\chi_{j}=\psi_{jx}$ are canonical momenta.

(ii) The coupled Neumann system

\begin{eqnarray}
\tilde{\xi}_{ixx} + \left( 2\sum_{j=0}^{g} b_{j}\tilde{\xi}_{j} \tilde{\eta}%
_{j} +\tilde{\xi}_{jx}\tilde{\eta}_{jx}\right)\tilde{\psi}_{i}= b_{i} \tilde{%
\xi}_{i} , \\
\tilde{\eta}_{ixx} + \left( 2\sum_{j=0}^{g} b_{j}\tilde{\xi}_{j} \tilde{\eta}%
_{j} +\tilde{\xi}_{jx}\tilde{\eta}_{jx}\right)\tilde{\eta}_{i}= b_{i} \tilde{%
\eta}_{i} .  \label{CouplNeum}
\end{eqnarray}
with constraint $\sum_{i=0}^{g}\tilde{\xi}_{i}\tilde{\eta}_{i} =1$, where $%
b_{0}< b_{1}< \ldots < b_{g}$ are fixed real numbers. The Neumann system is
obtained when $\xi_{i}=\eta_{i}$
\begin{eqnarray}
\tilde{\xi}_{ixx} + \left( 2\sum_{j=0}^{g} b_{j}\tilde{\xi}_{j}^2 +\tilde{\xi%
}_{jx}^2 \right)\tilde{\xi}_{i}= b_{i} \tilde{\xi}_{i} .  \label{Neum}
\end{eqnarray}
This system describes the motion of uncoupled harmonic oscillators $\tilde{%
\xi}_{ixx} = b_{i} \tilde{\xi}_{i}$, constrained by the force $%
\sum_{i=0}^{g}(b_{i}\tilde{\xi}_{i}^2 +\tilde{\xi}_{ix}^2)$ to move on the
unit sphere $\sum_{i=0}^{g}\tilde{\xi}_{i}^2=1$.
Let
\begin{equation}
\tilde{\xi}_{i}=\tilde{\psi}_{i}\exp(\tilde{\theta}_{i}),\,
\tilde{\eta}_{i}=\tilde{\psi}_{i}\exp(-\tilde{\theta}_{i}) ,\,
\sqrt{\tilde{f}}_{i} = \tilde{\psi}_{i}^{2}\tilde{\theta}_{ix} ,
\label{Deift1}
\end{equation}
then by Deift procedure (\ref{Deift1}), equations (\ref{CouplNeum}) transform
to Rosochatius system
\begin{eqnarray}
\tilde{r}_{ixx} = -\left( \sum_{j=0}^{g} b_{j}\tilde{r}_{j}^2 + \
\tilde{r}_{jx}^2- \frac{\tilde{f}_{j}}{\tilde{r}_{j}^{2}}
\right)\tilde{r}_{i}-\frac{\tilde{f}_{j}}{\tilde{r}_{j}^{3}}+
b_{i} \tilde{r}_{i} .  \label{Ros}
\end{eqnarray}
where $\sum_{i=0}^{g}\tilde{r}_{i}^2=1$.

\section{Baker-Akhiezer function}

\setcounter{equation}{0}

We review in this section some basic facts about Baker--Akhieser function
which will be used in the sequel.

Let $K$ be the hyperelliptic Riemann surface $\mu^2=\prod_{i=0}^{2g}
(\lambda-\lambda_{i})=R(\lambda ) $. The points of $K$ are pairs $%
P=(\lambda,R)$ and $\lambda(P)$ is the value of the natural projection $%
P\rightarrow\lambda(P)$ of $K$ to the complex projective line $CP^{1}$.

For given nonspecial divisor $D$, there is an unuque Baker-Akhiezer(BA)
function $\Psi(t,\lambda)$, such that

(i) the divisor of the poles of $\Psi$ is $D$,

(ii) $\Psi$ is meromorphic on $K\backslash\infty$

(iii) when $P\rightarrow \infty$
\begin{eqnarray}
\Psi(x,P)\exp(-kx) = 1 +\sum_{s=1}^{\infty} m_{s}(x) k^{-s} ,  \label{expan1}
\end{eqnarray}

is holomorphic and $k=\sqrt{\lambda(P)}$ is a local parameter near $P=\infty$%
.

There is a unique function $u(x)$ such that
\begin{equation}
\Psi_{xx} - u(x)\Psi= \lambda(P)\Psi ,  \label{Hill}
\end{equation}
where $\Psi$ is a BA function. Inserting expansion (\ref{expan1}) into (\ref
{Hill}), we obtain
\begin{eqnarray}
\Psi_{xx} - 2 m_{1x}(x)\Psi -\lambda(P)\Psi =\exp(kx) O(k^{-1}) ,
\end{eqnarray}
and due to the uniqueness of $\Psi$, we prove (\ref{Hill}), with $u(x)=2
m_{1x}(x)$.

By the Riemann-Roch theorem, there exists a unique differential $\tilde{%
\Omega}$ and a nonspecial divisor $D^{\tau}$ of degree $g$ such that the
zeros of $\tilde{\Omega}$ are $D + D^{\tau}$ and the expansion at $P=\infty$%
, $\tilde{\Omega}(P)=(1+O(k^{-2})) dk$.

For given nonspecial divisor $D^{\tau}$, there exists a unuque dual
Baker-Akhiezer (BA) function such that

(i) the divisor of the poles of $\Psi$ is $D^{\tau}$,

(ii) $\Psi$ is meromorphic on $K\backslash\infty$

(iii) when $P\rightarrow \infty$
\begin{eqnarray}
\Psi^{\tau}(x,P)\exp(-kx) = 1 +\sum_{s=1}^{\infty}\tilde{m}_{s}(x) k^{-s},
\label{expan}
\end{eqnarray}
Fix $\tau$ to be the hyperelliptic involution $P=(\lambda,R)\rightarrow(%
\lambda,-R)$, then we have $D^{\tau}=\tau D$, $\Psi^{\tau}(x,P)=\Psi(x,\tau
P)$. Let $\sum_{i=1}^{g}\mu_{i}(0)$ be the $\lambda$-projection of $D$, and $%
\sum_{i=1}^{g}\mu_{i}(x)$ be the $\lambda$-projection of the zero divisor of
$\Psi(x,P)$. The function $\Psi(t,P)\Psi^{\tau}(t,P)$ is meromorphic on $%
{\bf CP^{1}}$ and the following identity takes place
\begin{equation}
\Psi(x,P)\Psi^{\tau}(x,P)=\frac{F(x,\lambda)}{F(0,\lambda)}  \label{sqfun}
\end{equation}
where $F(x,\lambda)=\prod_{i=1}^{g}(\lambda - \mu_{i}(x))$. Introduce the
Wronskian
\begin{eqnarray}
\left\{\Psi(x,P),\Psi^{\tau}(x,P)\right\}=&&\Psi_{x}(x,P)\Psi^{\tau}(x,P)
-\Psi(x,P)\Psi_{x}^{\tau}(x,P) = \\
&&\frac{2\sqrt{R(\lambda)}}{\prod_{i=1}^{g}(\lambda -\mu_{i}(0)} ,  \nonumber
\end{eqnarray}
and the differential $\tilde{\Omega}$ is given explicitly by
\begin{equation}
\tilde{\Omega}(P)=\frac{1}{2}\prod_{i=1}^{g} (\lambda - \mu_{i}(0))/\sqrt{%
R(\lambda)}d\lambda .  \label{diff}
\end{equation}
We assume that $E(P)$ is a meromorphic function on $K$ with $g+1$ simple
poles $\infty,p_{1},\ldots,p_{g}$ and at $P\rightarrow\infty$, $%
E(P)=k+\ldots $, and $\tilde{E}(P)$ is meromorphic function with $g+1$
simple poles $q_{0},q_{1},\ldots ,q_{g}$ and at $P\rightarrow\infty$, $%
\tilde{E}(P)=k^{-1} +\ldots$. We also suppose that the divisors of poles of $%
E(P)$ and $\tilde{E}(P)$ are different from $D$, $D^{\tau}$.

Let
\begin{eqnarray}
\tilde{\xi}_{i}=\tilde{\xi}_{i}^{0}\Psi(x,q_{i}),\quad \tilde{\eta}_{i}=%
\tilde{\eta}_{i}^{0}\Psi^{\tau}(x,q_{i}),\quad  \nonumber \\
\tilde{\xi}^{0}_{i}\tilde{\eta}_{i}^{0}=Res_{P=q_i}E\tilde{\Omega}, \quad
b_{i}=\lambda(q_{i}),\,i=0,\ldots ,g  \label{solu1} \\
\xi_{i}=\xi_{i}^{0}\Psi(x,p_{i}),\quad
\eta_{i}=\eta_{i}^{0}\Psi^{\tau}(x,p_{i}),\quad  \nonumber \\
\xi^{0}_{i}\eta_{i}^{0}=Res_{P=q_i}\tilde{E}\tilde{\Omega}, \quad
a_{i}=\lambda(p_{i}),\,i=1,\ldots ,g  \label{solu2}
\end{eqnarray}
then
\begin{eqnarray}
& &u(x)=-\left(\sum_{i=0}^{g}b_{i}\tilde{\xi}_{i}\tilde{\eta}_{i} +\tilde{\xi%
}_{ix}\tilde{\eta}_{ix}\right),\quad \sum_{i=0}^{g}\tilde{\xi}_{i}\tilde{\eta%
}_{i}=1  \nonumber \\
& &u(x)=2\sum_{i=1}^{g}\xi_{i}\eta_{i}+ const.  \label{poten}
\end{eqnarray}
Let us construct the meromorphic differential $\tilde{E}\Psi\Psi^{\tau}%
\tilde{\Omega}$. By direct computations we have
\begin{equation}
\sum_{i=0}^{g}\,Res_{P=q_{i}}\,\tilde{E}\Psi\Psi^{\tau}\tilde{\Omega}
+Res_{P=\infty}\,\tilde{E}\Psi\Psi^{\tau}\tilde{\Omega} =\sum_{i=0}^{g}%
\tilde{\xi}_{i}\tilde{\eta}_{i} - 1 = 0,
\end{equation}
where $\tilde{\xi}_{i}^{0}\tilde{\eta}_{i}^{0}=Res_{P=q_{i}}\,\tilde{E}
\tilde{\Omega}$. Differentiating $\sum_{i=0}^{g}\tilde{\xi}_{i}\tilde{\eta}%
_{i} = 1$ twice and using Eq. (\ref{solu1}), we obtain first expression in (%
\ref{poten}). The eigenvalue equations
\begin{eqnarray}
&&\tilde{\xi}_{ixx}=(\lambda(q_{i}) + u(x))\tilde{\xi}_{i}, \\
&&\tilde{\eta}_{ixx}=(\lambda(q_{i}) + u(x))\tilde{\eta}_{i},
\end{eqnarray}
by replacing $u(x)$ from (\ref{poten}) are the coupled Neumann system (\ref
{CouplNeum}). By computations of the same kind, we have
\begin{equation}
\sum_{i=1}^{g}\,Res_{P=p_{i}}\,E\Psi\Psi^{\tau}\tilde{\Omega}
+Res_{P=\infty}\,E\Psi\Psi^{\tau}\tilde{\Omega} =\sum_{i=1}^{g} \xi_{i}
\eta_{i} - u(x)+\frac{1}{2} const. = 0,
\end{equation}
where $\xi_{i}^{0} \eta_{i}^{0}=Res_{P=p_{i}}\, E\tilde{\Omega}$. The
corresponding eigenvalue equations are the Garnier system (\ref{Garn}).

\subsection{Spectral interpretation}

Let $p$ be a positive real divisor of degree $g$ on a real hyperelliptic
curve
\begin{equation}
\mu^{2}=R(\lambda)=\prod_{i=0}^{2g}(\lambda-\lambda_{i}),\quad \lambda_{0} <
\lambda_{1} <,\ldots ,<\lambda_{2g} .
\end{equation}
The projection $\lambda(p_{i})$ lie in the closed lacunae $%
[\lambda_{2i-1},\lambda_{2i}]$. The following considerations, due to Jacobi,
can be used to construct such a divisor.

Each divisor $p$ determines and is determined by a system of polynomials
\begin{eqnarray}
&&\tilde{A}(\lambda)=\prod_{i=1}^{g}(\lambda-\lambda(p_{i})),\quad \tilde{C}%
(\lambda)=\tilde{A}(\lambda)\, \sum_{i=1}^{g}\,\frac{\sqrt{R(p_{i})}} {%
\tilde{A}^{\prime}(p_{i})(\lambda-\lambda(p_{i}))} ,  \nonumber \\
&&\tilde{B}(\lambda)=\lambda^{g+1}+\ldots ,
\end{eqnarray}
of degrees $g$, $g-1$, $g+1$, respectively, with $R=\tilde{C}^2-\tilde{A}%
\tilde{B}$. The complementary divisor $q$ is also determied by this
construction. This is the content of the following step.

For a given spectral data
\begin{equation}
\lambda_{0}=0 < \lambda_{1} < ,\ldots ,< \lambda_{2g},\quad
\lambda(p_{i})\in [\lambda_{2i-1},\lambda_{2i}],\,i=1,\ldots,g
\end{equation}
thete exists
\begin{equation}
\lambda(q_{i}),\,\,i=0,\ldots,g\quad\lambda(q_{0})\in
(-\infty,\lambda_{0}],\quad \lambda(q_{i})\in[\lambda_{2i-1},\lambda_{2i}] ,
\end{equation}
such that $R=\tilde{C}^2-\tilde{A}\tilde{B}$, and the projections $%
\lambda(q_{i})$ are the roots of $\tilde{B}$.

Note that the functions $E(P)$, $\tilde{E}(P)$ are meromorphic on $K$ and
the following formulas are immediate
\begin{equation}
E(P)=(\sqrt{R(\lambda)} +\tilde{C}(\lambda))/\tilde{A}(\lambda), \qquad
\tilde{C}(p_{i})=\sqrt{R(p_{i})}  \label{E}
\end{equation}

\begin{equation}
\tilde{E}(P)=(\sqrt{R(\lambda)} +\tilde{C}(\lambda))/\tilde{B}(\lambda),
\qquad \tilde{C}(q_{i})=-\sqrt{R(q_{i})}  \label{tE}
\end{equation}

Now we recall some facts from the periodic theory of Hill's equation. We
suppose that $u(x)$ is a real finite-gap potential, i.e. the operator $L$
has only $2g+1$ simple eigenvalues $\lambda_{0} < \lambda_{1} < ,\ldots
,<\lambda_{2g}$ and the rest of the spectrum consists of double eigenvalues.
The periodic spectra of $L$ is determined by the combined eigenvalues of the
periodic
\begin{equation}
L\,f_{2i} = \lambda_{2i}\,f_{2i},\quad f(x+1)=f(x),\,i=0,\ldots,g
\end{equation}
and the antiperiodic
\begin{equation}
L\,f_{2i-1} = \lambda_{2i-1}\,f_{2i-1},\quad f(x+1)=-f(x),\,i=1,\ldots,g
\end{equation}
eigenvalue equations. The intervals $(-\infty,\lambda_{0}],[\lambda_{2i-1},%
\lambda_{2i}]$ are termed lacunae. The Floquet solutions(periodic BA
function) and the corresponding Floquet multipliers, are given by
\begin{eqnarray}
&&\Psi(x,\lambda) =\left[ F(x,\lambda)/F(0,\lambda)\right]^{1/2}\,
\exp\left(\int_{0}^{x}\,\sqrt{R(\lambda)}/F(x^{\prime},\lambda)dx^{\prime}%
\right) ,  \label{BA1} \\
&&\Psi(x+1,\lambda)=\rho_{+}(\lambda)\Psi(x,\lambda) , \\
&&\Psi^{\tau}(x,\lambda) =\left[ F(x,\lambda)/F(0,\lambda)\right]^{1/2}\,
\exp\left(-\int_{0}^{x}\,\sqrt{R(\lambda)}/F(x^{\prime},\lambda)dx^{\prime}%
\right) ,  \label{BA2} \\
&&\Psi^{\tau}(x+1,\lambda)=\rho_{-}(\lambda)\Psi^{\tau}(x,\lambda) , \qquad
\rho_{\pm}=\exp(\pm\tilde{p}(\lambda)) ,
\end{eqnarray}
where
\begin{equation}
\tilde{p}(\lambda)=\int_{0}^{1}\,\sqrt{R(\lambda)}/F(x,\lambda)dx
\end{equation}
Note that if $\lambda$ is in the periodic spectrum, $\Psi(x,%
\lambda_{2i})=f_{2i}$, $i=0,\ldots,g$ is a periodic eigenfunction, and $%
\Psi(x,\lambda_{2i-1})=f_{2i-1}$,$i=1,\ldots ,g$ is an antiperiodic
eigenfunction. It is well known that the projections of the zeros of the
Floquet solution define the auxiliary spectrum of $L$.

The following expressions hold
\begin{eqnarray}
&& \xi_{i} \eta_{i}=\prod_{j=1}^{g}\,(\lambda(p_{i})-\mu_{j}(x))/\tilde{A}%
^{\prime}(p_{i}),  \label{Sol1} \\
&& \xi_{i}^{0} \eta_{i}^{0}=\prod_{j=1}^{g}\,(\lambda(p_{i})-\mu_{j}(0))/
\tilde{A}^{\prime}(p_{i}),  \nonumber \\
&& \tilde{\xi}_{i}\tilde{\eta}_{i}=\prod_{j=1}^{g}\,(\lambda(q_{i})-%
\mu_{j}(x))/ \tilde{B}^{\prime}(q_{i}),  \label{Sol2} \\
&& \tilde{\xi}_{i}^{0} \tilde{\eta}_{i}^{0}=\prod_{j=1}^{g}\,(%
\lambda(q_{i})-\mu_{j}(0))/ \tilde{B}^{\prime}(q_{i}).  \nonumber
\end{eqnarray}
Using (\ref{E}),(\ref{tE}) we obtain
\begin{eqnarray}
Res_{P=p_{i}}\,E(P)\Psi\Psi^{\tau}\tilde{\Omega}& & =
x_{i}^{0}y_{i}^{0}\Psi(x,p_{i})\Psi^{\tau}(x,p_{i})  \nonumber \\
& &=\prod_{j=1}^{g}\,(\lambda(p_{i})-\mu_{j}(x))/A^{\prime}(p_{i}) ,
\nonumber
\end{eqnarray}
where $\xi_{i}^{0}\eta_{i}^{0}$ is given by (\ref{Sol1}).

Let
\begin{eqnarray}
e_{2i}^2=\frac{\prod_{i\neq j}^{g}\,(\lambda_{2i} -\lambda_{2j})}
{\prod_{j\neq j=0}^{g}\,(\lambda_{2i}-\mu_{j}(0))} ,\quad f_{2i}^2=\frac{%
\prod_{j=1}^{g}\,(\lambda_{2i} -\mu_{j}(t))} {\prod_{i\neq
j=0}^{g}\,(\lambda_{2i}-\mu_{j}(0)) } = \Psi^2(\lambda_{2i}) ,
\end{eqnarray}
$i=0,\ldots ,g$, then the expressions (\ref{poten}), (\ref{Sol2}) are the
famous McKean-Moerbeke expansion of the potential $u(x)$ in terms of squares
of the eigenfunctions
\begin{equation}
u(x)=-\left( 2\sum_{i=0}^{g}\,\lambda_{2i} f_{2i}^2/e_{2i}
+\sum_{i=1}^{g}\lambda_{2i-1}-\sum_{i=1}^{g}\lambda_{2i} +\lambda_{0}\right),
\end{equation}
where the following identity among the squares of eigenfunctions hold on
\begin{equation}
\sum_{i=0}^{g}\,e_{2i}^{-2} f_{2i}^{2} =1.
\end{equation}

The results of this section, may be summarized by following:

Let $u(x)$ be a real nonsingular finite-gap potential. There exists $g$
eigenfunctions $\Psi(p_{1}),\ldots ,\Psi(p_{g})$ and $g+1$ eigenfunctions $%
\Psi(q_{0}),\ldots ,\Psi(q_{g})$ of Hill's equation, corresponding to the
eigenvalues $\lambda(p_{1}),\ldots,\lambda(p_{g})$ and $\lambda(q_{0}),%
\ldots ,\lambda(q_{g})$, respectively, such that

(i)
\begin{eqnarray}
u(x) = 2\sum_{i=1}^{g}\,\Psi(p_{i})\Psi^{\tau}(p_{i}) e_{i}^{-2} +
2\sum_{i=1}^{g}\lambda(p_{i})-\sum_{i=0}^{2g}\lambda_{i} ,  \\
e_{i}^{-2}=\prod_{j=1}^{g}(\lambda(p_{i})-\mu_{j}(0))/\tilde{A}%
^{\prime}\quad \Psi(p_{i})\equiv\Psi(x,\lambda)|_{\lambda =p_{i}} ,\,\,
i=1,\ldots,g . \nonumber
\end{eqnarray}

(ii)
\begin{eqnarray}
u(x) = 2\sum_{i=0}^{g}\,\lambda(q_{i})\Psi(q_{i})\Psi^{\tau}(q_{i})
e_{i}^{-2} - 2\sum_{i=0}^{g}\lambda(q_{i})+\sum_{i=0}^{2g}\lambda_{i} ,
\label{Expan1} \\
\tilde{e}_{i}^{-2}=\prod_{j=1}^{g}(\lambda(q_{i})-\mu_{j}(0))/\tilde{B}%
^{\prime}, \,\, \Psi(q_{i})\equiv\Psi(t,\lambda)|_{\lambda =q_{i}} ,\,\,
i=0,\ldots,g ,  \nonumber \\
\sum_{i=0}^{g}\tilde{e}^{-2}\Psi(q_{i})\Psi^{\tau}(q_{i}) =1 .
\end{eqnarray}
The corresponding eigenvalue equations are the Garnier and coupled Neumann
system.

Let
\begin{eqnarray}
& &e_{2i-1}^{2}=\prod_{i\neq j}^{g}(\lambda_{2i-1} - \lambda_{2j-1})/
\prod_{j=1}^{g}(\lambda_{2i-1} -\mu_{j}(0)) ,  \nonumber \\
& &f_{2i-1}^{2}=\prod_{j=1}^{g}(\lambda_{2i-1} - \mu_{j}(x))/
\prod_{j=1}^{g}(\lambda_{2i-1} -\mu_{j}(0)) ,\, i=1,\ldots,g  \nonumber
\end{eqnarray}
then we have the following expansion of the potential $u(x)$ in terms of
squares of antiperiodic eigenfunctions
\begin{equation}
u(x)=2\sum_{i=1}^{g}\,f_{2i-1}^2 e_{2i-1}^{-2} +2
\sum_{i=1}^{g}\lambda_{2i-1} -\sum_{i=0}^{2g}\lambda_{i} .
\end{equation}
We call the dynamical systems such in (i) , (ii), complementary dynamical
systems.

\subsection{Solutions in terms of auxiliary spectrum of Hill's equation}

The solutions of the Garnier system in terms of auxiliary spectrum $%
\mu_{j}(x)$, $j=1,\ldots,g$ are
\begin{eqnarray}
&&\xi_{i} =\xi_{i}^{0}[ F(x,a_{i})/F(0,a_{i})]^{1/2}\,
\exp\left(\int_{0}^{x}\,\sqrt{R(a_{i})}/F(x^{\prime},a_{i})dx^{\prime}%
\right) ,  \label{soluG1} \\
&&\eta_{i}=\eta_{i}^{0}[ F(x,a_{i})/F(0,a_{i}]^{1/2}\,
\exp\left(-\int_{0}^{x}\,\sqrt{R(a_{i})}/F(x^{\prime},a_{i})dx^{\prime}%
\right) ,  \label{soluG2}
\end{eqnarray}
where $\mu_{j}(x)$ satisfies the following system of differential equations
\begin{equation}
\frac{d}{dx}\mu_{j}(x) = 2\sqrt{R(\mu_{j})}/\prod_{j\neq k}^{g}(\mu_{j}(x) -
\mu_{k}(x)) ,  \label{muSys}
\end{equation}
with initial conditions
\begin{equation}
\mu_{j}(0)\in [\lambda_{2i-1},\lambda_{2i}],\quad
\xi_{i}^{0}\eta_{i}^{0}=F(0,a_{i})/\prod_{i\neq j}^{g}(a_{i}-a_{j}).
\end{equation}
Differentiating expressions
\begin{eqnarray}
\xi_{i}\eta_{i}=\xi_{i}^{0}\eta_{i}^{0}\Psi(x,p_{i})\Psi^{\tau}(x,p_{i}) =
\prod_{j=1}^{g}(\lambda(p_{i})-\mu_{j}(x))/\tilde{A}^{\prime}(p_{i}) ,
\end{eqnarray}
and
\begin{equation}
\xi_{ix}\eta_{i}-\xi_{i}\eta_{ix} =
\left\{\Psi(x,p_{i}),\Psi^{\tau}(x,p_{i})\right\}\xi_{i}^{0}\eta_{i}^{0} ,
\end{equation}
we have
\begin{eqnarray}
\Upsilon(x,P)|_{\lambda=\lambda(p_{i})} = \frac{d}{dx} \log \xi_{i}(t)
\nonumber \\
=[\frac{1}{2}\frac{d}{dx}\,\prod_{j=1}^{g}\,(\lambda -\mu_{j}(x)) +\sqrt{%
R(\lambda)}]/\prod_{j=1}^{g}\,(\lambda -\mu_{j}(x))
|_{\lambda=\lambda(p_{i})}  \label{chi1}
\end{eqnarray}
\begin{eqnarray}
\Upsilon^{\tau}(x,P)|_{\lambda=\lambda(p_{i})} = \frac{d}{dx} \log
\eta_{i}(t)  \nonumber \\
=[\frac{1}{2}\frac{d}{dx}\,\prod_{j=1}^{g}\,(\lambda -\mu_{j}(x)) -\sqrt{%
R(\lambda)}]/\prod_{j=1}^{g}\,(\lambda -\mu_{j}(x))
|_{\lambda=\lambda(p_{i})}  \label{chi2}
\end{eqnarray}
direct integration of (\ref{chi1}), (\ref{chi2}) gives the solutions (\ref
{soluG1}), (\ref{soluG2}). The function $\Upsilon(x,P)$ has $g$ poles at $%
\mu_{j}(x)$, then the numerator of (\ref{chi1}) is zero when $%
\lambda=\mu_{j}(x)$ and the following system takes place
\begin{equation}
\frac{d}{dx}\left(\prod_{j=1}^{g}\,(\lambda -\mu_{j}(x))\right)
|_{\lambda=\mu_{j}(x)} = 2\sqrt{R( \mu_{j}(x)) } .
\end{equation}
This is another form of the system (\ref{muSys}). In the same way, we can
obtain the solutions of the coupled Neumann system by replacing $%
\lambda(p_{i})$, $i=1,\ldots,g$ with $\lambda(q_{i})$, $i=0,\ldots,g$ in (%
\ref{soluG1}), (\ref{soluG2}).

Let $\lambda(p_{i})$ be the antiperiodic eigenvalues $\lambda_{2i-1}$, $%
i=1,\ldots,g$. Then the exponential function in (\ref{soluG1}), (\ref{soluG2}%
) cancel, $\xi_{i}^{0}=\eta_{i}^{0}$ and the solutions of the $g$%
-dimensional oscillator are
\begin{equation}
\xi_{i}^{2}=\prod_{j=1}^{g}\,(\lambda_{2i-1} - \mu_{j}(x))/ \prod_{i\neq
j}^{g}\,(\lambda_{2i-1}-\lambda_{2j-1}) .
\end{equation}

Let $\lambda(p_{i})=a_{i}$ be in a general position, i.e. $\lambda(p_{i})\in
[\lambda_{2i-1},\lambda_{2i}] $ and by Deift elimination procedure we may
identify $\xi_{i}$ with $\xi_{i}^{0}[F(x,a_{i})/F(0,a_{i})]^{1/2}$ and
\begin{equation}
\theta_{i}=\int_{0}^{x}\,\sqrt{R(a_{i})}/\prod_{i=1}^{g}(a_{i}-\mu_{j}(x^{%
\prime}))dx^{\prime}, \quad \xi_{i}^{0}=\eta_{i}^{0} ,
\end{equation}
and, hence, the solutions of the Rosochatius-Wojciechowski system are
\begin{equation}
\xi_{i}^2=\prod_{j=1}^{g}\,(a_{i}-\mu_{j}(x))/ \prod_{i\neq
j}^{g}(a_{i}-a_{j}) .  \label{GarnT}
\end{equation}
Inserting explicit expression of BA-function given by (\ref{BA1}) in Hill's
equation we have
\begin{eqnarray}
\frac{1}{2} F_{xx}(x,\lambda)F(x,\lambda) - \frac{1}{4} F_{x}^{2}(x,%
\lambda)-(u(x)+\lambda)F^{2}(x,\lambda)= -R(\lambda) .  \label{qEQ}
\end{eqnarray}
The polynomial solution of (\ref{qEQ}) below we will call Novikov polynomial
\cite{Nov}. Assuming that Novikov polynomial dependts on time $t$,
the zero curvature representation for KdV hierarchy of equations
have the following form
\begin{equation}
M_{t}(\lambda')-L_x(\lambda^{\prime})+\left[M(\lambda`),L(\lambda^{\prime})
\right ] , \label{ZCeq}
\end{equation}
where matrices $L$ and $M$ are given by
\begin{eqnarray}
L(\lambda^{\prime})&=&\left(
\begin{array}{cc}
-F_{x}(x,t,\lambda^{\prime})/2 & F(x,t,\lambda^{\prime}) \\
-F_{xx}(x,t,\lambda^{\prime})/2 +Q(x,t,\lambda')F(x,t,\lambda') &
 F_{x}(x,t,\lambda^{\prime})/2
\end{array}
\right)  \nonumber \\
M^{\prime}(\lambda^{\prime})&=&\left(
\begin{array}{cc}
0          &   1 \\
Q(x,t,\lambda') &   0
\end{array}
\right)\,.  \nonumber
\end{eqnarray}
The equation (\ref{ZCeq}) is equivalent to
\begin{equation}
\frac{\partial Q}{\partial t}= -2\left[\frac{1}{4}\partial^3_x
-Q(x,t,\lambda^{\prime})\partial_x -\frac{1}{2} Q_{x}(x,t,\lambda^{\prime})
\,\right]\cdot F(x,\lambda^{\prime})\,.  \label{Geq}
\end{equation}
where $Q(x,t,\lambda^{\prime})=u(x)+\lambda^{\prime}$
in the case of KdV hierarchy. Equation (\ref{Geq}) is called the generating
equation. For a different choices of the form of $F(x,t,\lambda^{\prime})$ and
$Q(x,t,\lambda^{\prime})$, this procedure leads to different hierarchies of
integrable equations, as an example to the KdV, nonlinear Shr\"{o}dinger and
sine-Gordon hierarchies or to the Dym hierarchy \cite{ACHM}.
The Lax representation $L_{x}=[M,L]$ yields the hyperelliptic curve
$K=(\mu^{\prime},\lambda^{\prime})$
\begin{eqnarray}
&&\mbox{Det}(L(\lambda^{\prime})-\mu^{\prime}\mbox{I})=0 , \\
&&\mu^{2}=-\frac{1}{2}F F_{xx}+\frac{1}{4}F_{x}^{2}+(\lambda'+u)F=R(\lambda') ,
\nonumber
\end{eqnarray}
generating the integrals of motion for stationary KdV hierarchy.

Using the equation (\ref{qEQ}) and the following expansion of potential
$u(x)$
in terms of squares of eigenvalue functions $\xi_{k}^{2}(x)=\beta_{k}(x)$
\begin{equation}
u(x)=2\sum_{i=1}^{g}\xi_{k}^2(x)+2\sum_{i=1}^{g} a_{i}
-\sum_{k=0}^{2g}\lambda_{i} ,
\end{equation}
have the form
\begin{eqnarray}
\frac{1}{2}\beta_{kxx}\beta_{k}-\frac{1}{4}\beta_{k}^{2}-
(u(x)+a_{k})\beta_{k}^{2}=-\frac{R(a_{k})}{a_{kj}}=-f_{k}    \label{SolGarnT}
\end{eqnarray}
where we use the solutions $F(x,a_{k})/a_{kj}$ of Rosochatius-Wojciechowski
system and $a_{kj}\equiv\prod_{k\neq j}^{g}(a_{k}-a_{j})$. Denoting $%
f_{k}=R(a_{k})/a_{kj}$ and $d=\sum_{i=1}^{g}a_{i}-\frac{1}{2}%
\sum_{k=0}^{2g}\lambda_{k}$ for the original variable $\beta_{k}=\xi_{k}^{2}$
we have the following equation
\begin{eqnarray}
\xi_{kxx}=2\left(\sum_{i=1}^{g}\xi_{k}^{2} +d\right)\xi_{k}-\frac{f_{k}}{%
\xi_{k}^{3}} .
\end{eqnarray}

To understand the role of GD polynomials and of their generating function
in the construction of a map between stationary and restricted
flows of KdV equation and exact solution of completely integrable systems
related to Hill's equation let us consider the following system:
\begin{eqnarray}
&&p_{00}-v_{1}=a_{0}, \quad P_{0}\left(v_{1}-\sum_{j=1}^{n}\beta_{j}\right),
\\
&&P^{\lambda_{k}}\beta_{k}=0, \qquad (k=1,\ldots,n)
\end{eqnarray}
where$\lambda_{1},\ldots,\lambda_{n}$ are fixed parameters, $%
P^{\lambda_{k}}:=P_{1}+4\lambda_{k} P_{0}$ ($P_{0}$ and $P_{1}$ being the
two KdV Poisson operators). This is a system of $(g+2)$ equations in $u,
v_{1},\beta_{1},\ldots,\beta_{g}$. The second equation will be referred to
as the $P_{0}$-restriction of the first KdV flow $X_{0}=P_{0}v_{1}=v_{1x}$,
and the last $n$ equations define the kernel of $g$ Poisson operators
extracted from the Poisson operator. On account of (\ref{GD}), (\ref{Opers})
and (\ref{Brel}) this system is equivalent to the following one:
\begin{eqnarray}
u=\frac{v_{1}}{2}+\frac{a_{0}}{4},\quad v_{1}=\sum_{j=1}^{n}\beta_{j} +
c,\quad B^{\lambda_{k}}(\beta_{k},\beta_{k})=f_{k} ,
\end{eqnarray}
where $c$ and $f_{k}$ are free parameters and $B^{\lambda}$ is just the
generating function of the GD polynomials.

Using the first two equations to eliminate $u$ and $v_{1}$ from the last $g$
equations, one gets a system of $n$ second-order ODEs for $%
\beta_{1},\ldots,\beta_{g}$ :
\begin{eqnarray}
2\beta_{kxx}\beta_{k}-\beta_{kx}^{2}+2\beta_{k}^{2}
\left(\sum_{j=1}^{n}\beta_{j} +d\right) -\lambda_{k}\beta_{k}^{2} = f_{k},
\quad k=1,\ldots,g  \label{QRos}
\end{eqnarray}
where $d:=c+a_{0}/2$. Introducing the so-called eigenfunction variables $%
\psi_{j}^{2}=\beta_{j}$ and the momenta $\chi_{j}=\psi_{jx}$, equations (\ref
{QRos}) can be written in canonical Hamiltonian form
\begin{eqnarray}
\psi_{jx}=\frac{\partial K_{G}}{\partial \chi_{j}}, \quad \chi_{jx}=-\frac{%
\partial K_{G}}{\partial \psi_{j}}, \quad j=1,\ldots,n
\end{eqnarray}
with Hamiltonian
\begin{eqnarray}
K_{G}=\sum_{j=1}^{n}\chi_{j}^{2}
-\left(\sum_{k=1}^{n}\psi_{j}^{2}\right)^{2}- \sum_{j=1}^{n}
a_{j}\psi_{j}^{2}- \sum_{j=1}^{n}\frac{f_{j}}{\psi_{j}^{2}} .
\end{eqnarray}
A set of integrals of motion is
\begin{eqnarray}
I_{j}=&&\chi_{j}^{2}-\psi_{j}^{2}\left(a_{j}+
\sum_{k=1}^{g}\psi_{k}^{2}\right)- \frac{f_{j}}{\psi_{j}^{2}}+  \nonumber \\
&& \sum_{k\neq j}^{n}\frac{1}{a_{jk}} \left(-\frac{f_{j}\psi_{k}^{2}}{%
\psi_{j}^{2}} - \frac{f_{k}\psi_{j}^{2}}{\psi_{k}^{2}}+
(\psi_{j}\chi_{k}-\psi_{k}\chi_{j})^{2}\right) .  \label{IntRosI}
\end{eqnarray}
where we denote $a_{jk}=a_{j}-a_{k}$.

Now we shall construct a map between the $g$-th
stationary flow and the previous restricted flow of the KdV hierarchy. To
this end we extend the corresponding phase spaces, regarding some free
parameters in the Hamiltonian functions as addiitional dynamical variables.
As for the $P_{1}$-formulation of the stationary flow we extend its phase
space to a $(3g+1)$-dimensional space, $\tilde{M}_{n}$, with coordinates $%
(q_{k},p_{k};a_{0},\ldots,a_{g-1},a_{2g})$; analogously we consider the $%
P_{0}$-formulation of the first restricted flow in the extended space $%
\tilde{{\it M}}_{g}$ with coordinates $(\psi_{k},\chi_{k};f_{1},%
\ldots,f_{k},d)$.

Let us consider the solutions $q_{k}$ of the dynamical equations (\ref{Deq}%
); then $v^{(g)}(\lambda)$ given by
\begin{equation}
v^{(g)}(\lambda)=\lambda(q^{2}(\lambda))^{(g-1)}-q_{g}^{2} ,
\end{equation}
with $q(\lambda)=1+\sum_{j=1}^{g}\,q_{j}\lambda^{-j}$, satisfies (2.17), and
consequently satisfies the following equation:
\begin{equation}
B^{\lambda}(v^{(g)},v^{(n)})=\lambda^{2g} \mbox{d}(\lambda) ,
\end{equation}
where, as above, we put $u=v_{1}/2+a_{0}/4$. So, for each $g$-tuple of
distinct complex parameters $a_{j}$, any solution $v^{(g)}(\lambda)$
fulfills the system
\begin{equation}
B^{a_{k}}(v^{(g)}(a_{k}),v^{(g)}(a_{k})=a_{k}^{2g} %
\mbox{d}(a_{k}) , \quad (k=1,\ldots,g)
\end{equation}
where $v^{(g)}(a_{k}):=v^{(\lambda)}|_{\lambda=a_{k}}$. In order to have
a solution also satisfying the second equation $v_{1}=
\sum_{j=1}^{g}\beta_{j} +c$, the Lagrange interpolation formula can be used.
It allows us to represent the polynomial $v^{n}(\lambda)$ by
\begin{equation}
v^{(n)}(\lambda)=a(\lambda)\left(1+\sum_{j=1}^{g}\frac{\beta_{j} }
{\lambda-a_{j} }\right) ,  \label{LagIF}
\end{equation}
where $a(\lambda)=\prod_{j=1}^{g}(\lambda-a_{j})$, and
\begin{equation}
\beta_{j}=\frac{v^{(g)}(a_{k})}{a^{\prime}(a_{k})},
(k=1,\ldots,g) .
\end{equation}
($a^{\prime}(\lambda)$ means the derivative of $a(\lambda)$ with respect to $%
\lambda$.

Obviously the $g$ functions $\beta_{k}$ are solutions of the following
system
\begin{eqnarray}
2\beta_{kxx}\beta_{k}-\beta_{kx}^{2}+2\beta_{k}^{2}
\left(\sum_{j=1}^{g}\beta_{j} +d\right) -\lambda_{k}\beta_{k}^{2} = \frac{%
\lambda_{k}^{2g}\mbox{d}(a_{k})}{(a^{\prime}(a_{k}))^{2}} , \quad
k=1,\ldots,g  \label{QRos1}
\end{eqnarray}
Furthermore, $\beta_{k}$ satisfy the so-called Bargmann constraint
\begin{equation}
\sum_{j=1}^{g} (\beta_{j}-a_{j}) = v_{1} ,
\end{equation}
as one can verify by means of (\ref{LagIF}).

The function $B^{\lambda}$ is also a generating function of integrals of
motion for Garnier system. Indeed evaluating the function $B^{\lambda}$ by
means of (\ref{LagIF}) and eliminating the first $x$-derivatives of $%
\chi_{k} $ by means of Hamilton equations (\ref{Deq}), one gets
\begin{eqnarray}
4\sum_{j=1}^{g}\frac{I_{j}}{\lambda-\lambda_{j}} + \sum_{j=1}^{g} \frac{f_{j}%
}{(\lambda-\lambda_{j})^{2}} + 2d-\lambda=\frac{\lambda^{2g}\hat{a}(\lambda)%
}{(a(\lambda))^{2}} ,
\end{eqnarray}
where $I_{j}$ are the functions. Taking in this equation the residues at $%
\lambda=a_{j}$ it follows that the functions $I_{j}$ are integrals of motion
along the flow (\ref{Deq}).

Let $\lambda(q_{i})=b_{i}$ be in a general position, i.e.

$\lambda(q_{i})=b_{i}\in (-\infty,\lambda_{0}],
[\lambda_{2i-1},\lambda_{2i}] $ and by Deift elimination procedure we may
identify $\tilde{\psi}_{i}$ with $\tilde{\xi}%
_{i}^{0}[F(t,b_{i})/F(0,b_{i})]^{1/2}$ and
\begin{equation}
\tilde{\theta}_{i}=\int_{0}^{x}\,\sqrt{R(b_{i})}/\prod_{i=1}^{g}
(b_{i}-\mu_{j}(x^{\prime}))dx^{\prime},
\end{equation}
and, hence, the solutions of the Rosochatius system are
\begin{equation}
\tilde{\psi}_{i}^2=\prod_{j=0}^{g}\,(b_{i}-\mu_{j}(x))/ \prod_{i\neq
j}^{g}(b_{i}-b_{j}) .
\end{equation}
where $i,j=0,\ldots g$. Now we illustrate the general aproach with some
simple examples

{\bf Example 1} Let $u(x)=6\wp(x+\omega^{\prime})$ be the two-gap Lam\'e
potential with simple periodic spectrum (see for example \cite{EK})
\begin{equation}
\lambda_{0}=-\sqrt{3g_{2}},\quad \lambda_{1}=-3e_{0}\quad ,
\lambda_{2}=-3e_{1}\quad , \lambda_{3}=-3e_{2}\quad , \lambda_{4}=\sqrt{%
3g_{2}}.
\end{equation}
and the corresponding Hermite polynomial have the form
\begin{equation}
F(\wp(x+\omega^{\prime}),\lambda)=\lambda^{2}-
3\wp(x+\omega^{\prime})\lambda+
9\wp^{2}(x+\omega^{\prime})-\frac{9}{4}g_{2} .  \label{HerPol}
\end{equation}
Consider the following genus $2$ nonlinear anisotropic oscillator with
Hamiltonian
\begin{equation}
H=\frac{1}{2}(p_{1}^{2}+p_{2}^{2})+\frac{1}{4}(q_{1}^{2}+q_{2}^{2})^{2}-
\frac{1}{2}(a_{1}q_{1}^{2}+a_{2}q_{2}^{2}),
\end{equation}
where $(q_{i},p_{i})$, $i=1,2$ are canonical variables with $p_{i}=q_{ix}$
and $a_{1},a_{2}$ are arbitrary constants. The
simple solutions of these system are given in terms of Hermite polynomial
\begin{equation}
q_1^2=2\frac{F(x,\tilde{\lambda}_{1})} {\tilde{\lambda}_{2}-\tilde{\lambda}%
_{1}} ,\quad q_2^2=2\frac{F(x,\tilde{\lambda}_{2})} {\tilde{\lambda}_{1}-%
\tilde{\lambda}_{2}} ,
\end{equation}
Let us list the corresponding solutions

\begin{itemize}
\item  Periodic solutions expressed in terms of single Jacobian elliptic
functions
\end{itemize}

The nonlinear anisotropic oscillator admit the following solutions:
\begin{eqnarray}
q_1 & = & C_1 \mbox{sn}(\alpha x, k), \\
q_2 & = & C_2 \mbox{cn}(\alpha x, k),  \label{onegap}
\end{eqnarray}
where amplitudes $C_1$ ,$C_2$ and temporal pulsewidth $1/\alpha$ of are
defined by parameters $a_{1}$ and $a_{2}$ as
\begin{eqnarray}
\alpha^2 k^2 & = & a_{2}-a_{1}  \nonumber \\
C^{2}_{1} & = & a_{2} + \alpha^2 - \alpha^2 k^2 ,  \nonumber \\
C^{2}_{2} & = & a_{1} + \alpha^{2} ,  \nonumber
\end{eqnarray}
where $0 < k < 1$.

Following our spectral method it is clear, that the solutions \ref{onegap}
are associated with eigenvalues $\lambda_2 = - e_2$ and $\lambda_3 = - e_3$
of one -- gap Lam\'e potential.

\begin{itemize}
\item  Periodic solutions expressed in terms of products of Jacobian
elliptic functions
\end{itemize}

\begin{eqnarray}
q_1 & = & C \mbox{dn}(\alpha x, k) \mbox{sn}(\alpha x, k), \\
q_2 & = & C \mbox{dn}(\alpha x ,k ) \mbox{cn}(\alpha x, k),  \label{Flor}
\end{eqnarray}
where $\mbox{sn}$,$\mbox{cn}$, $\mbox{dn}$ are the standard Jacobian
elliptic functions \cite{BE}, $k$ is the modulus of the elliptic functions $%
0 < k < 1$, an the wave characteristic parameters: amplitude $C$, temporal
pulsewidth $1/\alpha$ and $k$ are related to the physical parameters and, $k$
through the following dispersion relations
\begin{eqnarray}
C^{2} & = & \frac{2 (4a_{2} - a_{1})}{5} ,  \nonumber \\
k^{2} & = & \frac{(4a_{2}-a_{1})} {15} ,  \nonumber \\
\alpha^{2} & = & \frac {5 (a_{2}-a_{1})} {4a_{2}-a_{1}} .  \nonumber
\end{eqnarray}
We have found the following solutions of the nonlinear oscillator
\begin{eqnarray}
q_1 & = & C\alpha^2 k^2 \mbox{cn}(\alpha x,k)\mbox{sn}(\alpha x,k) \\
q_2 & = & C\alpha^2 \mbox{dn}^2 (\alpha x, k) + C_{1}  \label{UzKos}
\end{eqnarray}
where $C$, $C_1$, $\alpha$ and $k$ are expressed through parameters
$a_{1}$ and $a_{2}$ by the following relations
\begin{eqnarray}
C^2 & = & \frac {18} {a_{2}-a_{1}} ,  \nonumber \\
C_1 & = & \frac { C (4a_{1}-a_{2})}{5},\quad
 \nonumber \\
k^2 & = & \frac {2 \sqrt { \frac{5}{3} (a_{2}^{2}-a_{1}^{2})}}
{2 \sqrt{\frac{5}{3} (a_{2}^{2}-a_{1}^{2})}+aa_{2}-3a_{1}},
\nonumber \\
\alpha^2 & = & \frac {1}{10} ( 2 a_{2}-3a_{1}+
\sqrt{\frac{5}{3}(a_{2}^{2}-a_{1}^2) } ).
\end{eqnarray}

\begin{itemize}
\item  Periodic solutions associated with the two-gap Treibich-Verdier
potentials
\end{itemize}

Below we construct the two periodic solutions associated with the
Treibich-Verdier potential. Let us consider the potential
\begin{equation}
u(x)=6\wp(x+\omega^{\prime})+2{\frac{(e_1-e_2)(e_1-e_3)}{\wp(x+\omega^{%
\prime})-e_1}}  \label{tv4}
\end{equation}
and construct the solution in terms of Lam\'e polynomials associated with
the eigenvalues $\tilde{\lambda}_1,\tilde{\lambda}_2$, $\tilde{\lambda}_1 >
\tilde{\lambda}_2$
\begin{eqnarray}
\tilde{\lambda}_1&=&e_2+2e_1+2\sqrt{(e_1-e_2)(7e_1+2e_2)},  \nonumber \\
\tilde{\lambda}_2&=&e_3+2e_1+2\sqrt{(e_1-e_3)(7e_1+2e_3)}.  \label{zz}
\end{eqnarray}
The finite and real solutions $q_1,q_2$ have the form
\begin{eqnarray}
q_1&=& C_{1}\mbox{sn}(z,k)\mbox{dn}(z,k) +C_{2}\mbox{sd}(z,k) ,  \nonumber \\
q_2&=& C_{3}\mbox{cn}(z,k)\mbox{dn}(z,k) +C_{4}\mbox{cd}(z,k),  \label{ee2}
\end{eqnarray}
where $C_{i}$, $i=1,\ldots 4$ are constants and have important geometrical
interpretation \cite{EK}. The concrete expressions in terms of $k,\tilde{%
\lambda}_{1},\tilde{\lambda_{2}}$ are given in \cite{CEEK}

Analogously we can find the elliptic solution associated with the
eigenvalues
\begin{eqnarray}
\tilde{\lambda}_1&=&e_2+2e_1+2\sqrt{(e_1-e_2)(7e_1+2e_2)},\quad \tilde{%
\lambda}_2=-6e_1,  \label{zz2}
\end{eqnarray}
We have
\begin{eqnarray}
q_1&=&\tilde{C}_{1}\mbox{dn}^{2}(z,k) \\
q_2&=& C_{1}\mbox{sn}(z,k)\mbox{dn}(z,k) +C_{2}\mbox{sd}(z,k) ,  \label{ee3}
\end{eqnarray}
where $C$ is given in \cite{CEEK}.

The general formula for elliptic solutions of genus $2$ nonlinear
anisotropic oscillator is given in \cite{CEEK}
\begin{eqnarray}
q_1^2&=&{\frac{1}{\tilde{\lambda}_2-\tilde{\lambda}_1}}
\left(2\tilde{\lambda}_1^2+2\tilde{\lambda}_1\sum_{i=1}^N
\wp(x-x_i) \right. \nonumber
\\ &&+\left.6\sum_{1\leq i< j\leq N}\wp(x-x_i)\wp(x-x_j)-{\frac{Ng_2}{4}}+
\sum_{1\leq i< j\leq 5}\lambda_i\lambda_j\right),  \label{q1} \\
q_2^2&=&{\frac{1}{\tilde{\lambda}_1-\tilde{\lambda}_2}}
\left(2\tilde{\lambda}_2^2+2\tilde{\lambda}_2\sum_{i=1}^N
\wp(x-x_i) \right. \nonumber
\\ &&+\left.6\sum_{1\leq i< j\leq N}\wp(x-x_i)\wp(x-x_j)-{\frac{Ng_2}{4}}+
\sum_{1\leq i< j\leq 5}\lambda_i\lambda_j\right),  \label{q2}
\end{eqnarray}
where $x_{i}$ are solutions of equations $\sum_{i\neq j}\wp'(x_{i}-x_{j})=0,
j=1,\ldots, N$ and $N$ is positive integer.

{\bf Example 2}  Garnier type system.
Consider the following genus $2$ Garnier type system with
Hamiltonian
\begin{equation}
H=\frac{1}{2}(p_{1}^{2}+p_{2}^{2})+\frac{1}{4}(q_{1}^{2}+q_{2}^{2})^{2}-
\frac{1}{2}(\gamma_{1} q_{1}^{2}+\gamma_{2} q_{2}^{2})+\frac{f_{1}}{q_{1}^{2}}+
\frac{f_{2}}{q_{2}^{2}},
\end{equation}
where $(q_{i},p_{i})$, $i=1,2$ are canonical variables with $p_{i}=q_{ix}$
and $a_{1},a_{2}$ are arbitrary constants. The
simple solutions of these system are given in terms of Hermite polynomial
in the following form (\ref{GarnT})
\begin{equation}
q_1^2=2\frac{F(x,a_{1})}
{a_{1}-a_{2}} , \quad
q_2^2=2\frac{F(x,a_{2})}
{a_{2}-a_{1}} ,
\end{equation}
the same settings of periodic spectra, Lam\'e potential and Hermite
polynomials (\ref{HerPol})
as in example 2. The main results from the general theory are the following:

    i) the parameters $f_{1}$ and $f_{2}$ are expressed in terms of algebraic
curve by (\ref{SolGarnT})
\begin{eqnarray}
f_{1}=\frac{R(a_{1})}{a_{1}-a_{2}}, \quad
f_{2}=\frac{R(a_{2})}{a_{2}-a_{1}}, \nonumber \\
R(\lambda)=(\lambda^{2}-3g_{2})(\lambda+3e_{1})(\lambda+3e_{2})(\lambda+3e_{3})
. \nonumber
\end{eqnarray}

    ii) the parameters
$a_{1}$ and $a_{2}$ must lie in one or other of the intervals
\begin{equation}
[-\sqrt{3g_{2}}, 3e_{1}], [3e_{2},3e_{3}], [\sqrt{3g_{2}},\infty)
\end{equation}
and $\gamma_{1}=3a_{1}+2a_{2}, \gamma_{2}=2a_{1}+3a_{2}$.

These results are in complete agreement with solutions obtained by different
method in recent paper \cite{pp98}.

{\bf Example 3} Simple solutions of the H\'enon-Heiles type system.

We consider a generalized H\'enon-Heiles type system with two-degrees of
freedom.

Its Hamiltonian is
\begin{equation}
H_{0}=\frac{1}{2}(p_{1}^{2}+p_{2}^{2}) + q_{1}^{3}+\frac{1}{2}q_{1}
q_{2}^{2} + \frac{a_{4}}{8 q_{2}^{2}} +\frac{a_{0}}{2}\left(q_{1}^{2} +\frac{%
1}{4} q_{2}^{2}\right) -\frac{a_{1}}{4} q_{1} ,  \label{HH}
\end{equation}
where $q_{1}, q_{2}, p_{1}, p_{2}$ are the canonical coordinates and momenta
and $a_{0}, a_{1}, a_{4}$ are free constant parameters. This Hamiltonian
encompasses the two cases $a_{0}=a_{4}=0$ and $a_{0}=a_{1}=0$ introduced in
\cite{18}. Moreover $H_{0}$ is related with the Hamiltonian
\begin{equation}
H_{H}= \frac{1}{2}(p_{1}^{2}+p_{2}^{2}) + \bar{q}_{1}^{3}+\frac{1}{2}\bar{q}%
_{1} \bar{q}_{2}^{2} + \frac{\bar{a_{4}}}{{8 \bar{q}_{2}^{2}}}
+\frac{1}{2}\left(A\bar{q}_{1}^{2} + B \bar{q}_{2}^{2}\right) ,
\end{equation}
through the map
\begin{equation}
q_{1}=\bar{q}_{1} + \frac{A}{2} -2B,\quad q_{2}=\bar{q}_{2}, \quad a_{0}=-2
A+12 B, \quad a_{1}=-A^{2}+16 AB-48 B^{2} .
\end{equation}
The function $H_{H}$ is the Hamiltonian of a classical integrable
H\'enon-Heiles system with the additional term $a_{4}/8\bar{q}%
_{2}^{2}$.

The function (\ref{HH}) is the Hamiltonian of the vector field obtained
reducing $X_{0}(u)=u_{x}$ to the stationary manifold $M_{2}$ given by the
fixed points of the flow $X_{2}+c_{1}X_{1}+c_{2}X_{0}$
\begin{equation}
M_{2}={u|u^{(5)}+10u_{xxx}u+20u_{xx}u_{x}+30u_{x}u^{2}+
c_{1}(u_{xxx}+6u_{x}u) +c_{2}u_{x}=0}
\end{equation}
where $c_{1}=-a_{0}/2$, $c_{2}=-a_{1}/2+a_{0}^{2}/4$.

It can be obtained specializing to the case $g=2$ the Hamiltonian of the $%
P_{1}$-system. In this case $H_{2}^{(1)}=H_{0}$ and the canonical
coordinates and momenta are, respectively, $q_{1}=v_{1}/2$, $%
q_{2}^{2}=-v_{2} $, $p_{1}=q_{1x}$, $p_{2}=q_{2x}$. The integrals of motion
obtained by the reduction of the GD polynomials are
\begin{eqnarray}
&&H_{0}\equiv -\frac{1}{8}p_{02}|_{x} \\
&&H_{2}\equiv -\frac{1}{8}p_{22}|_{x} =-\frac{a_{4}}{8} \\
&&H_{1}\equiv -\frac{1}{8}p_{12}|_{x}=
\end{eqnarray}
where $-\frac{1}{8}p_{12}|_{x}$ is given by
\begin{eqnarray}
p_{2}q_{1}-p_{1}p_{2}q_{2}-\frac{1}{2}q_{1}^{2}q_{2}^{2}-\frac{1}{8}%
q_{2}^{4}+ \frac{a_{4}q_{1}}{4q_{2}^{2}}-\frac{a_{0}}{4}q_{1}q_{2}^{2}+
\frac{a_{1}}{8}q_{2}^{2} .
\end{eqnarray}
Next we will derive $(2\times 2)$ matrix Lax representation for generalized
H\'enon-Heiles system (\ref{HH}). Using Lax representation $L_{x}=[M,L]$,
particular case of eq. (\ref{ZCeq}) i.e. when there is no time dependence, we
have
\begin{eqnarray}
&&F(x,\lambda)=\lambda^2+\frac{1}{2} q_{1}\lambda-\frac{1}{16}q_{2}^{2}, \quad
\quad
V=-F_{x}/2=-\frac{1}{4}p_{1}\lambda + \frac{1}{16} q_{2} p_{2} , \nonumber \\
&&W=-F_{xx}/2+Q F=\lambda^{3}-(\frac{1}{2}q_{1}+\frac{1}{4} a_{0})\lambda^2+
\nonumber \\
&&(\frac{1}{4} q_{1}^{2}+\frac{1}{16} q_{2}^{2}-\frac{1}{16}a_{1}+
\frac{1}{8} a_{0} q_{1})\lambda + \frac{1}{16}p_{2}^{2} +
\frac{1}{64} \frac{a_{4}}{q_{2}^{2}} ,
\quad Q(x,\lambda)=\lambda-q_{1}-\frac{1}{4}a_{0} . \nonumber
\end{eqnarray}
The corresponding algebraic curve have the form
\begin{eqnarray}
\mu^2=\lambda^{5}-\frac{1}{4}a_{0}\lambda^{4}-\frac{1}{16}a_{1}\lambda^{3}
+8 H_{0}\lambda^{2}+32\,H_{1}\lambda-\frac{1}{1024}a_{4} .
\end{eqnarray}
Using explicit expression for Hermite polinomial (\ref{HerPol}) we obtain
the following simple solutions for the system (\ref{HH}):
\begin{eqnarray}
q_{1}=-6\wp(x+\omega'), \qquad
q_{2}^2=-16\,(9\wp(x+\omega')^{2} -\frac{9}{4} g_{2}) .
\end{eqnarray}
where $a_{0}=0, a_{1}=3.4.7 g_{2}, A_{4}=-3^{4}.4^{4} g_{2} g_{3}$.

\section{$2\times 2$ Lax representation and $r$-matrix approach}

\setcounter{equation}{0} The Lax equation for completely integrable systems
discussed in the previous section
\begin{equation}
L_x(\lambda^{\prime})=\left[M(\lambda^{\prime}),L(\lambda^{\prime})\right]%
\,, \label{Lax}
\end{equation}
with matrices $L$ and $M$ given by
\begin{eqnarray}
L(\lambda^{\prime})&=&\left(
\begin{array}{cc}
V(x,\lambda^{\prime}) & U(x,\lambda^{\prime}) \\
W(x,\lambda^{\prime}) & -V(x,\lambda^{\prime})
\end{array}
\right)  \label{Llm1} \\
M(\lambda^{\prime})&=&\left(
\begin{array}{cc}
0 & 1 \\
Q(x,\lambda^{\prime}) & 0
\end{array}
\right)\,.  \label{Llm2}
\end{eqnarray}
is equivalent to the Garnier system, where $U(x,\lambda^{\prime}),
V(x,\lambda^{\prime}), W(x,\lambda^{\prime}), Q(x,\lambda^{\prime})$ have
the form
\begin{eqnarray}
U(x,\lambda^{\prime})=a(\lambda^{\prime})\left(1-\sum_{i=1}^{g} \frac{%
\xi_{i}\eta_{i}}{\lambda^{\prime}- a_{i}}\right), \,
V(x,\lambda^{\prime})=-\frac{1}{2}U_{x}(x,\lambda^{\prime}) \\
W(x,\lambda^{\prime})=a(\lambda^{\prime})\left(\lambda^{\prime}+%
\sum_{i=1}^{g}\,\xi_{i}\eta_{i}+ \sum_{i=1}^{g}\frac{\xi_{ix}\eta_{ix}}{%
\lambda^{\prime}- a_{i}}\right), \quad
Q(x,\lambda^{\prime})=\lambda^{\prime}+ 2\sum_{i=1}^{g}\xi_{i}\eta_{i} .
\end{eqnarray}
Finally we point out one usefull expression, which is easy to derive from
Lax representation (\ref{Lax})
\begin{equation}
W(x,\lambda^{\prime})=U(x,\lambda^{\prime})Q(x,\lambda^{\prime})-\frac{1}{2}
U(x,\lambda^{\prime})_{xx}.  \label{W}
\end{equation}
The Lax representation yields the hyperelliptic curve $K=(\mu^{\prime},%
\lambda^{\prime})$
\begin{equation}
\mbox{Det}(L(\lambda^{\prime})-\mu^{\prime}\mbox{I})=0 ,
\end{equation}
generating the integrals of motion $H, F^{(i)}, i=1,\ldots,g$. We have
\begin{equation}
\mu^{2}=V^{2}(x,\lambda^{\prime})+U(x,\lambda^{\prime})
W(x,\lambda^{\prime}) ,  \label{Lcurve}
\end{equation}
From (\ref{Lcurve}) and explicit expressions of $U(x,\lambda^{\prime}),
V(x,\lambda^{\prime}), W(x,\lambda^{\prime})$ we obtain
\begin{equation}
\mu^{2}=a(\lambda^{\prime})^{2}\left(\lambda^{\prime}+ \sum_{i=1}^{g}\frac{%
H_{i}}{\lambda^{\prime}-a_{i}} + \frac{1}{4}\sum_{i=1}^{g}\frac{J_{i}^{2}}{%
(\lambda^{\prime}-a_{i})^{2}} +\frac{1}{2}\sum_{i=1}^{g}\frac{I_{i}}{%
\lambda^{\prime}- a_{i}}\right) ,
\end{equation}
where
\begin{eqnarray}
I_{i}&=&\sum_{k\neq i}\frac{
(\xi_{k}\eta_{ix}-\eta_{i}\xi_{kx}) (\eta_{k}\xi_{ix}-\xi_{i}\eta_{kx})}{%
a_{k}-a_{i}} ,  \nonumber \\
&+&\sum_{k\neq i}\frac{
(\xi_{i}\xi_{kx}-\xi_{k}\xi_{ix}) (\eta_{i}\eta_{kx}-\eta_{k}\eta_{ix})}{%
a_{k}-a_{i}} ,  \nonumber \\
&&H_{i}=\xi_{ix}\eta_{ix}-a_{i}\xi_{i}\eta_{i}-\xi_{i}\eta_{i}
\left(\sum_{k=1}^{g}\xi_{k}\eta_{k}\right) , \,
J_{i}=\xi_{ix}\eta_{i}-\xi_{i}\eta_{ix},
\end{eqnarray}
and $\sum_{i=1}^{g}H_{i}$ is the Hamiltonian for Garnier system. Simple
reduction $\eta_{i}= \xi_{i}^{*}$ gives us the second flow of stationary
vector nonlinear Schr\"{o}dinger equation, where by $*$ we denote complex
conjugation. The complementary to the last system is complex Neumann system
(see for example \cite{IA})
\begin{eqnarray}
\tilde{\xi}_{ixx}+2\left(\sum_{j=0}^{g} b_{j}|\tilde{\xi}_{j}|^{2} +|\tilde{%
\xi}_{jx}|^{2}\right)\tilde{\xi}_{i}=b_{i}\tilde{\xi}_{i} .
\end{eqnarray}
with $\sum_{i=0}^{g}|\xi_{i}|^{2}=1$.
Using the Deift elimination procedure we obtain new $2\times 2$ Lax pair for
Rosochatius-Wojciechowski system. Bellow we list only the final results for
Lax pair elements of considered in this paper dynamical systems:

\begin{itemize}

\item Rosochatius-Wojciechowski system

\begin{eqnarray}
&&U(x,\lambda^{\prime})=a(\lambda^{\prime})\left(1-\sum_{i=1}^{g} \frac{%
\psi_{i}^{2}}{\lambda^{\prime}- a_{i}}\right), \,
V(x,\lambda^{\prime})=-\frac{1}{2}U_{x}(x,\lambda^{\prime}) \nonumber \\
&&W(x,\lambda^{\prime})=a(\lambda^{\prime})\left(\lambda^{\prime}+%
\sum_{i=1}^{g}\,\psi_{i}^{2} +
\sum_{i=1}^{g}\frac{1}{%
\lambda^{\prime}- a_{i}}
(\psi_{ix}^{2}-\frac{f_{i}}{\psi_{i}^{2}})\right) ,
\\ &&
Q(x,\lambda^{\prime})=\lambda^{\prime}+ 2\sum_{i=1}^{g}\psi_{i}^{2} .
\nonumber
\end{eqnarray}

\item Rosochatius system

\begin{eqnarray}
&&U(x,\lambda^{\prime})=b(\lambda^{\prime})\left(\sum_{i=1}^{g} \frac{%
\psi_{i}^{2}}{\lambda^{\prime}- b_{i}}\right), \,
V(x,\lambda^{\prime})=-\frac{1}{2}U_{x}(x,\lambda^{\prime}) \nonumber \\
&&W(x,\lambda^{\prime})=b(\lambda^{\prime})\left(1-%
+ \sum_{i=1}^{g}\frac{1}{%
\lambda^{\prime}- b_{i}}
(\psi_{ix}^{2}-\frac{f_{i}}{\psi_{i}^{2}})\right) ,
\\ &&
Q(x,\lambda^{\prime})=\lambda^{\prime}+ 2\sum_{i=1}^{g}\psi_{i}^{2} ,
\nonumber
\end{eqnarray}
where $b(\lambda^{\prime})=\prod_{i=0}^{g}(\lambda^{\prime}-b_{i})$.

\item second stationary flow of vector NLSE

\begin{eqnarray}
&&U(x,\lambda^{\prime})=a(\lambda^{\prime})\left(1-\sum_{i=1}^{g} \frac{%
|\xi_{i}|^{2}}{\lambda^{\prime}- a_{i}}\right), \,
V(x,\lambda^{\prime})=-\frac{1}{2}U_{x}(x,\lambda^{\prime})\nonumber \\
&&W(x,\lambda^{\prime})=a(\lambda^{\prime})\left(\lambda^{\prime}+%
\sum_{i=1}^{g}\,|\xi_{i}|^{2}+ \sum_{i=1}^{g}\frac{|\xi_{ix}|^{2}}{%
\lambda^{\prime}- a_{i}}\right),
\\ &&
Q(x,\lambda^{\prime})=\lambda^{\prime}+ 2\sum_{i=1}^{g}|\xi_{i}|^{2} .
\nonumber
\end{eqnarray}

\item complex Neumann system

\begin{eqnarray}
&&U(x,\lambda^{\prime})=b(\lambda^{\prime})\left(\sum_{i=1}^{g} \frac{%
|\xi_{i}|^{2}}{\lambda^{\prime}- b_{i}}\right), \,
V(x,\lambda^{\prime})=-\frac{1}{2}U_{x}(x,\lambda^{\prime})\nonumber \\
&&W(x,\lambda^{\prime})=a(\lambda^{\prime})\left(1-%
 \sum_{i=1}^{g}\frac{|\xi_{ix}|^{2}}{%
\lambda^{\prime}- b_{i}}\right),
\\ &&
Q(x,\lambda^{\prime})=\lambda^{\prime}+ 2\sum_{i=1}^{g}|\xi_{i}|^{2} .
\nonumber
\end{eqnarray}

\end{itemize}

Finally we want to point out that  $I_{i}$
for Rosochatius-Wojciechowski  system coincide with expression given in
(\ref{IntRosI}). Another Lax equation have the following form
\[
L_x(\lambda^{\prime})=\left[M(\lambda`),L(\lambda^{\prime})\right]\,,
\]
where matrices $L$ and $M$ are given by
\begin{eqnarray}
L(\lambda^{\prime})&=&\left(
\begin{array}{cc}
-F_{x}(x,\lambda^{\prime})/2 & F(x,\lambda^{\prime}) \\
-F_{xx}(x,\lambda^{\prime})/2 & F_{x}(x,\lambda^{\prime})/2
\end{array}
\right) \equiv  \nonumber \\
&& \left(
\begin{array}{cc}
V(x,\lambda^{\prime}) & U(x,\lambda^{\prime}) \\
W^{\prime}(x,\lambda^{\prime}) & -V(x,\lambda^{\prime})
\end{array}
\right) \, ,  \label{KdV1} \\
M^{\prime}(\lambda^{\prime})&=&\left(
\begin{array}{cc}
0 & 1 \\
0 & 0
\end{array}
\right)\,.  \label{KdV2}
\end{eqnarray}
where we made the following identification $U(x,\lambda^{\prime})=F(x,%
\lambda^{\prime})$. The Poisson bracket relations for the matrix $%
L(\lambda^{\prime})$ \cite{KRT,T1} are closed into the following $r$-matrix
algebra
\begin{equation}
\left\{{L_{1}}(\lambda^{\prime}),{L_{2}}(\mu^{\prime})\right\}=
[r_{12}(\lambda^{\prime},\mu^{\prime}),{L_{1}}(\lambda^{\prime})]-[r_{21}(%
\lambda^{\prime},\mu^{\prime}), {L_{2}}(\mu^{\prime})\,]\, .  \label{Lalg}
\end{equation}
Here the standard notations are introduced:
\begin{eqnarray}  \label{r1}
&&L_{1}(\lambda^{\prime})= L(\lambda^{\prime})\otimes I\,,\qquad
L_{2}(\mu^{\prime})=I\otimes L(\mu^{\prime})\,,  \nonumber \\
\\
&&r_{12}(\lambda,\mu)=\frac{\Pi}{\lambda-\mu}\,\qquad
r_{21}(\lambda,\mu)=\Pi r_{12}(\mu,\lambda)\Pi\,,  \label{r2}
\end{eqnarray}
and $\Pi$ is the permutation operator of auxiliary spaces.

The Poisson bracket relations for the Lax matrix $L(\lambda^{\prime})$ are
preassigned by the initial symplectic structure. It is necessary to
calculate only two brackets
\begin{equation}
\left\{F(x,\lambda^{\prime}),F(x,\mu^{\prime})\right\}=0\,,  \label{Br1}
\end{equation}
and
\begin{eqnarray}  \label{Br2}
\left\{V(x,\lambda^{\prime}),F(x,\mu^{\prime})\right\}&=&  \nonumber \\
&&\left\{ F(x,\lambda^{\prime})\sum_{j=1}^g \frac{\mbox{g}_{jj}\,p_j(x)}
{\lambda^{\prime}-\mu_{j}(x)}, \prod_{j=1}^g(\mu^{\prime}-\mu_{j}(x))\right\}
\nonumber \\
&=& -F(x,\lambda^{\prime})F(x,\mu^{\prime})\sum_{j=1}^n \frac{\mbox{g}_{jj}}{%
(\lambda^{\prime}-\mu_{j}(x))(\mu^{\prime}-\mu_{j}(x))} \nonumber \\
&=&\frac{F(x,\lambda^{\prime})F(x,\mu^{\prime})}{\lambda^{\prime}-\mu^{%
\prime}} \sum_{j=1}^g \left(\frac{\mbox{g}_{jj}}{\lambda^{\prime}-\mu_{j}(x)}%
- \frac{\mbox{g}_{jj}}{\mu^{\prime}-\mu^{\prime}_{j}(x)}\right)  \nonumber \\
&=&\frac{1}{\lambda^{\prime}-\mu^{\prime}}\left[F(x,\mu^{\prime})-
F(x,\lambda^{\prime})\right]\,,
\end{eqnarray}
where we used a standard decomposition of rational function
\[
F(x,\lambda^{\prime})^{-1}= \sum_{j=1}^n\frac{\mbox{g}_{jj}}{%
\lambda-\mu_{j}(x)}\,,\qquad \mbox{g}_{jj}=\left.\mbox{Res}%
\right|_{\lambda^{\prime}=\mu_{j}(x)} F(x,\lambda^{\prime})^{-1}(\lambda)\,.
\]
and the following definitions
\begin{eqnarray}
&&\mbox{g}_{jj}=\mbox{Res}_{|\lambda=\mu_{j}(x)}F^{-1}(x,\lambda)= \frac{1}{%
\prod_{k\neq j}(\mu_{j}(x)-\mu_{k}(x))}, \\
&&p_{j}(x)=V(x,\lambda)_{|\lambda=\mu_{j}(x)}=\sqrt{R(\mu_{j}(x))}.
\end{eqnarray}

Another Poisson brackets may be directly derived from these brackets and by
definition of the entries of the Lax matrix $L(\lambda)$ via derivative of
the single function $F(x,\lambda^{\prime})$
\begin{eqnarray}  \label{Br3}
\left\{V(x,\lambda^{\prime}),V(x,\mu^{\prime})\right\}&=&0\,  \nonumber \\
\left\{W(x,\lambda^{\prime}),F(x,\mu^{\prime})\right\}&=&\frac{d}{dx}
\left\{V(x,\lambda^{\prime}),F(x,\mu^{\prime})\right\}=  \nonumber \\
&&\frac{2}{\lambda^{\prime}-\mu^{\prime}}\left[V(x,\lambda^{\prime})-V(x,%
\mu^{\prime}) \right]\, ,  \nonumber \\
\\
\left\{W(x,\lambda^{\prime}),F(x,\mu^{\prime})\right\}&=&-\frac{1}{2}\frac{%
d^2}{dx^{2}} \left\{V(x,\lambda^{\prime}),F(x,\mu^{\prime})\right\}=
\nonumber \\
&&\frac{1}{\lambda^{\prime}-\mu^{\prime}} \left[W(x,\lambda^{\prime})-W(x,%
\mu^{\prime})\right]\, ,  \nonumber \\
\left\{W(x,\lambda^{\prime}),W(x,\mu^{\prime})\right\}&=&-\frac{1}{2}\frac{%
d^3}{dx^{3}} \left\{V(x,\lambda^{\prime}),F(x,\mu^{\prime})\right\}=0 \, .
\nonumber
\end{eqnarray}

Applying the following transformation directly to the Lax representation $%
L(\lambda^{\prime})$ we obtain a family of the new Lax pairs \cite{KRT,T1}
\begin{eqnarray}  \label{LnLax}
&&L^{\prime}(\lambda^{\prime})=L(\lambda^{\prime})-\sigma_-\cdot
\left[\phi(x,\lambda^{\prime})F(x,\lambda^{\prime})^{-1}\right]_{N}\,,
\qquad \sigma_-=\left(
\begin{array}{cc}
0 & 0 \\
1 & 0
\end{array}
\right)\,,  \nonumber \\
\\
&&M^{\prime}(\lambda')=M-\sigma_{-}\cdot\left[\phi(x,\lambda')F(x,\lambda')^{-2
} \right]_{N} = \left( \begin{array}{cc}
0 & 1 \\
Q(x,\lambda') & 0
\end{array}
\right)\,.  \nonumber
\end{eqnarray}
Here $\phi(x,\lambda^{\prime})$ is a function on spectral parameter and
$[z]_{N}$ means restriction of $z$ onto the ${\rm ad}^*_R$-invariant
Poisson subspace of the initial $r$-bracket. For the rational $r$-matrix we
can use the linear combinations of the following Taylor projections
\begin{equation}
{[ z ]_{N}}=\left[\sum_{k=-\infty}^{+\infty} z_k\lambda^k\,
\right]_{N}\equiv \sum_{k=0}^{N} z_k\lambda^k\,,  \label{Cut}
\end{equation}
or the Laurent projections.

New Lax matrix $L^{\prime}(\lambda^{\prime})$ \cite{KRT,T1} obeys the linear
$r$-bracket, where constant $r_{ij}$-matrices substituted by $%
r_{ij}^{\prime} $-matrices depending on dynamical variables.
\begin{eqnarray}
&&r_{12}(\lambda^{\prime},\mu^{\prime})\rightarrow r^{\prime}_{12}=r_{12}
\nonumber \\
&& -\frac{\left( [\phi(\lambda)
F(x,\lambda^{\prime})^{-2}(\lambda^{\prime})]_{N}- [\phi(\mu)
F(x,\mu')^{-2}]_{N}\,\right)} {(\lambda^{\prime}-
\mu^{\prime})}\cdot\sigma_{-}\otimes\sigma_{-} \,.  \label{dpr}
\end{eqnarray}

\section{Conclusions}

In this paper we have given new exact solutions of the physically
significant completely integrable dynamical systems. These solutions can be
interpreted as eigenfunctions of suitable differential operator. New example
of complementary dynamical systems (stationary second flow of the vector
nonlinear Schr\"odinger equation and complex Neumann system) is presented.

\vspace{2cm} {\bf\Large Acknowledgements} \\*\\*I wish to thank to Prof.
V. Gerdjikov, who poined out me the papers \cite{7,8,9}, book \cite{v98} and
for valuable discussions.


\begin{thebibliography}{99}
\bibitem{Mos1}  Moser J 1980 Geometry of quadrics and spectral theory, Chern
Symposium 1979 Berkeley (Berlin: Springer)

\bibitem{20}  Moser J 1980 Various aspects of integrable Hamiltonian systems
{\it Dynamical Systems}, CIME 1978 (Progress in Mathematics 8)
(Basel:Birkh\"auser) 233-89

\bibitem{Mos2}  Moser J 1981 Integrable hamiltonian systems and spectral
theory, Pisa: Lezioni Fermiane

\bibitem{Knor}  Kn\"{o}rer H 1982 Geodesic on quadrics and a mechanical
problem of Neumann, J. Reine Angew. Math. {\bf 334} 67-78

\bibitem{Ves}  Veselov A~P 1980 Finite-band potentials and an integrable
system on the sphere with quadratic potential, Funct. Anal. Appl. {\bf 14}
48-50

\bibitem{5}  Cao C~W 1990 Non linearization of eigenvalue problem {\it
Non Linear Physics} ed. C~Gu et al (Berlin: Springer)

\bibitem{2}  Antonowicz M and Rauch-Wojciechowski S 1992 How to construct
finite dimensional bi-Hamiltonian systems from soliton equations: Jacobi
integrable potentials J. Math. Phys. {\bf 33} 2115-25

\bibitem{F}  Flaschka H 1983 Relations between infinite - dimensional and
finite - dimensional isospectral equations{\it Non-linear Integrable
Systems-Classical and Quantum Theory} (Singapore: World Scientific)

\bibitem{S}  Schilling R~J 1987 Generalizations of the Neumann system-A
curve theoretical approach, {\it Comm. Pure Appl. Math.} {\bf 40}, 455

\bibitem{AHP}  Adams M~R, Harnad~J, and Previato~E 1988 Isospectral
Hamiltonian flows in finite and infinite dimensions I. Generalized Moser
systems and Moment maps into loop algebras, {\it Commun. Math. Phys. }
{\bf 117} 451-500

\bibitem{K}  Kostov N~A 1989 Quasi-periodic solutions of the integrable
dynamical systems related to Hill's equation, {\it Lett. Math. Phys.}
{\bf 17} 95-108

\bibitem{AHH}  Adams M~R, Harnad~J, and Hurtubise~J 1990 Isospectral
Hamiltonian flows in finite and infinite dimensions II. Integration of
flows, {\it Commun. Math. Phys.} {\bf 134} 555-585

\bibitem{AHH1}  Adams M~R, Harnad~J, and Hurtubise~J 1993 Darboux
coordinates and Liouville-Arnold integration in loop algebras, Commun. Math.
Phys. {\bf 155} 385-413

\bibitem{FT}  Florjanzyk~M and Tremblay~R Periodic and solitary waves in
bimodal optical fibers, Phys. Lett. {\bf A 25} 34-36

\bibitem{KU}  Kostov~N~A and Uzunov~M 1992 New kinds of periodical waves in
birefringent optical fibers, Opt. Commun. {\bf 89} 389-92

\bibitem{EK}  Enol'skii V~Z and Kostov N~A 1994 On the geometry of elliptic
solitons, {\it Acta Applicandae Mathematicae} {\bf 36} 57-86

\bibitem{CEEK}  Christiansen P~L, Eilbeck J~C, Enolskii V~Z and Kostov N~A
1995 Quasi-periodic solutions of the coupled nonlinear Schr\"odinger
equations {\it Proc. R. Soc. Lond. A} {\bf 451}, 685-700

\bibitem{BBEIM}  Belokolos E~D, Bobenko~A~I, Enolskii~V~Z, Its~A~R and
Matveev~V~B 1994, Algebraical-geometrical methods in the theory of
integrable equations (Berlin: Springer)

\bibitem{GW95a} Gesztesy~F, Weikard~R, 1995
{\it Lam\'e potentials and the stationary
(m)KdV hierarchy}, Math. Nachr. {\bf 176} , 73--91.

\bibitem{GW95b} Gesztesy~F, Weikard~R, 1995
{\it Treibich-Verdier potentials and the stationary (m)KdV
hierarchy}, Math. Z. {\bf 219} , 451--476.

\bibitem{GW95c} Gesztesy~F, Weikard~R, 1995
{\it On Picard potentials}, Diff. Int. Eqs. {\bf 8}
, 1453--1476.

\bibitem{GW95d} Gesztesy~F, Weikard~R, 1995
{\it A characterization of elliptic
finite-gap
potentials}, C. R. Acad. Sci. Paris {\bf 321} , 837--841.

\bibitem{GW96}  Gesztesy~F, Weikard~R, 1996
{\it Picard potentials and Hill's equation on a torus},
Acta Math. {\bf 176}, 73--107.

\bibitem{GW98}  Gesztesy~F, Weikard~R, 1998
{\it A characterization of all elliptic algebro-geometric
solutions of the AKNS hierarchy}, Acta Math. {\bf 181} , to
appear.





\bibitem{Nov}  Novikov S~P 1974 Periodic problem for KdV equation I. Funct.
Anal. Appl. {\bf 8} 54-66

\bibitem{KM}  McKean H~P and van Moerbecke, 1975 The spectrum of Hill's
equation, Invent. Math. {\bf 30}, 217-74

\bibitem{DMN}  Dubrovin B~A, Matveev V~B and Novikov S~P Nonlinear equations
of KdV type, finite-zone operators and abelian varieties, Russ. Math. Surv.
{\bf 31}, 59-146

\bibitem{KT}  McKean H~P and Trubowitz, 1976 Hill's operator and
hyperelliptic function theory in the presence of infinitely many branch
points, Commun. Pure Appl. Math {\bf 29}, 143-26

\bibitem{Kr}  Krichever I 1977 The method of algebraic geometry in the
theory of nonlinear equations, Uspekhi. Mat. Nauk. {\bf 32} 183-208

\bibitem{Cher}  Cherednik I~M 1978 Differential equations for the
Baker-Akhiezer functions of algebraic curves Functs. Anal. Prilozh. {\bf
12} 45-54

\bibitem{AM80} Adler~M, van Moerbecke~P, 1980 {\it Completely integrable
systems, euclidean Lie algebras, and curves}, Adv. in Math. {\bf 38} 267-317



\bibitem{HPM}  McKean H~P 1985 Variations on a theme of Jacobi, Commun. Pure
and Appl. Math. {\bf XXXVIII} 669-78

\bibitem{F1}  Flaschka H 1984 Towards an algebro-geometric interpretation of
the Neumann system, Tohoku Math. Journ. {\bf 36} 407-26

\bibitem{Hor}  Horozov E~I 1984 On exact solutions of Neumann's problem of
moving point on the sphere with quadratic potential, Compt. Rend. Acad.
Bulg. Sci. {\bf 37} 145-148 (in russian)

\bibitem{M}  Mumford D 1984 Tata Lectures on Theta, vol. 2
(Basel,Boston,Stuttgard: Birkhauser Verlag)

\bibitem{Skl}  Sklyanin E~K 1985 Separation of variables: New trends, Progr.
Theor. Phys. Suppl. {\bf 118} 35-60

\bibitem{1}  Dickey L~A 1991 Soliton Equations and Hamiltonian Systems
(Singapore: World Scientific)

\bibitem{3}  Fordy A~P 1991 The H\'enon-Heiles system revisited Physica
{\bf 52D} 204-10

\bibitem{4}  Antonowicz M and Rauch-Wojciechowski S 1992 Bi-Hamiltonian
Formulation of the H\'enon-Heiles System and its Multidimensional Extensions
{\it Phys. Lett.} {\bf 163A} 167-72

\bibitem{6}  Antonowicz M and Rauch-Wojciechiwski S 1990 Constrained flows
and integrable PDEs and bi-Hamiltonian structure of the Garnier system
{\it Phys. Lett.} {\bf 147A} 455-62

\bibitem{7}  Tondo G 1994 A connection between the H\'enon-Heiles system and
the Garnier system {\it Theor. Math. Phys.} {\bf 33} 796-802 (1994
Teor. Mat. Fiz {\bf 99} 552-9)

\bibitem{8}  Tondo G 1995 On the integrability of stationary and restricted
flows of the KdV hierarchy {\it J. Phys. A: Math. Gen.} {\bf 28}
5097-5115

\bibitem{9}  Morosi C and Tondo G 1997 Quasi-bi-Hamiltonian systems and
separability {\it J. Phys A: Math. Gen.} {\bf 30} 2799-2806


\bibitem{10}  Caboz R, Ravoson V and Gavrilov L 1991 Bi-Hamiltonian
structure of an integrable H\'enon-Heiles system {\it J. Phys. A: Math.
Gen.} {\bf 24} L523-5

\bibitem{11}  Magri F 1978 A simple model of the integrable Hamiltonian
equation {\it J. Math. Phys.} {\bf 19} 1156-62


\bibitem{v98} Vilasi G 1998 {\it Hamiltonian dynamics} (World Scientific,
Singapore)

\bibitem{CC}  Choodnovsky D~V and Choodnovsky G~V~ 1978 {\it Lett. Nuovo
Cimento} {\bf 22}, 47

\bibitem{W}  Wojciechowski S 1985 Integrability of one particle in a
perturbed central quartic potential {\it Phys. Scr.} {\bf 31} 433-8

\bibitem{13}  Marsden J~E and Ratiu T 1986 Reduction of Poisson manifolds
{\it Lett. Math. Phys.} {\bf 11} 161-9

\bibitem{18}  Blaszak M and Rauch-Wojciechowski S 1994 A H\'enon-Heiles
system and related integrable Newton equations, J. Math. Phys. {\bf 35}
1693-709


\bibitem{Al}  Alber S~I 1981 On Stationary Problems for Equations of
Korteweg-de Vries Type {\it Commun. Pure Appl. Math.} {\bf 34} 259-72

\bibitem{AM}  Alber M~S and Marsden J~E 1992 On geometric phases for soliton
equations {\it Commun. Math. Phys.} {\bf 149}, 217-40

\bibitem{ACHM}  Alber M~S, Camassa R, Holm D~D, and Marsden J~E 1995 The
geometry of peaked solitons and billiard solutions of class of integrable
PDE's {\it Lett. Math. Phys.} {\bf 32}, 137-51

\bibitem{BE}  Bateman~H and Erdelyi~A 1955 {\it Higher transcedental
functions} vol. 2 (New York:McGraw-Hill)

\bibitem{Z}  Zakharov V~E 1998 Description of the $n$-orthogonal curvilinear
coordinate systems and hamiltonian integrable systems of the hydrodynamic
type. Part I. Integration of the Lam\'{e} equations, {\it Duke
Math. Journ.} {\bf 94}, 103-139


\bibitem{Krich1}  Krichever I~M 1996 Algebraic-geometrical $n$-orthogonal
curvilinear coordinate systems and solutions to associativity equations
{\it hep-th/9611158}

\bibitem{KKM}  Kalnins E~G, Kuznetsov V~B, and Miller Jr. 1994 Quadrics on
complex Riemannian spaces of constant curvature, separation of variables and
Gaudin magnet {\it J. Math. Phys.} {\bf 35} 1710-31


\bibitem{pp98} Porubov A~V, Parker D~F 1998 Some general solutions to
coupled nonlinear Schr\"odinger equations, {\it Wave motion} {\bf 900}
1-13


\bibitem{KRT}  Kulish P.P, Rauch-Wojciechowski, and Tsiganov A~V 1996
Stationary problems for equations of the KdV type and dynamical $r$-matrices
{\it J. Math. Phys.} {\bf 37} 3365

\bibitem{EEKT}  Eilbeck J~C, Enolskii V~Z, Kuznetsov V~B, and Tsiganov A~V
1994 Linear $r$-matrix algebra for classical separable systems {\it J.
Phys. A} {\bf 27} 567-78

\bibitem{T1}  Tsiganov A~V 1998 The St\"{a}ckel systems and algebraic curves
Teor. Math. Phys. {\bf 115}, 3-28 (in russian) {\it solv-int/9712003}

\bibitem{T2}  Tsiganov A~V 1996 Authomorphisms of $sl(2)$ and dynamical $r$%
-matrices solv-int/9610003; 1998 J. Math. Phys. {\bf 39} 650-664

\bibitem{IA}  Ii~K and Arira~Y, 1987 A completely integrable Hamiltonian
system of C. Neumann-type on the complex space, Tokyo J. Math. {\bf 10}
77-86



\end{thebibliography}
\end{document}